# Simulating Customer Experience and Word-Of-Mouth in Retail - A Case Study

**Peer-Olaf Siebers**
**Uwe Aickelin**
University of Nottingham, School of Computer Science
Nottingham, NG8 1BB, UK
*pos@cs.nott.ac.uk*

**Helen Celia**
**Chris W. Clegg**
University of Leeds, Centre for Organisational Strategy, Learning & Change (LUBS)
Leeds, LS2 9JT, UK

Agents offer a new and exciting way of understanding the world of work. In this paper we describe the development of agent-based simulation models, designed to help to understand the relationship between people management practices and retail performance. We report on the current development of our simulation models which includes new features concerning the evolution of customers over time. To test the features we have conducted a series of experiments dealing with customer pool sizes, standard and noise reduction modes, and the spread of customers' word of mouth. To validate and evaluate our model, we introduce new performance measure specific to retail operations. We show that by varying different parameters in our model we can simulate a range of customer experiences leading to significant differences in performance measures. Ultimately, we are interested in better understanding the impact of changes in staff behavior due to changes in store management practices. Our multi-disciplinary research team draws upon expertise from work psychologists and computer scientists. Despite the fact we are working within a relatively novel and complex domain, it is clear that intelligent agents offer potential for fostering sustainable organizational capabilities in the future.

**1. Introduction**

The retail sector has been identified as one of the biggest contributors to the productivity gap, whereby the productivity of the UK lags behind that of France, Germany and the USA [1, 2]. A recent report into UK productivity asserted that, '...the key to productivity remains what happens inside the firm and this is something of a 'black box'...' [3]. A recent literature review of management practices and organizational productivity concluded that management practices are multidimensional constructs that generally do not demonstrate a straightforward relationship with productivity variables [4], and that both management practices and productivity measures must be context specific to be (respectively) effective and meaningful. Many attempts have been made to link management practices to an organization's productivity and performance (for a review, see [5]), however research findings have been so far been mixed, creating the opportunity for the application of different techniques to advance our understanding.

Simulation can be used to analyze the operation of dynamic and stochastic systems showing their development over time. There are many different types of simulation, each of which has its specific field of application. Agent-Based Simulation (ABS) is particularly useful when complex interactions between system entities exist such as autonomous decision making or proactive behavior. Agent-Based Modeling (ABM) shows how micro-level processes affect macro level outcomes; macro level behavior is not explicitly modeled, it emerges from the micro-decisions made by the individual entities [6].

There has been a fair amount of modeling and simulation of operational management practices, but people management practices have often been neglected although research suggests that they crucially impact upon an organization's performance [7]. One reason for this relates to the key component of people management practices, an organization's people, who may often be unpredictable in their individual behavior. Previous research into retail productivity has typically focused on consumer behavior and efficiency evaluation (e.g. [8, 9]), and we seek to build on this work and address the neglected area of people management practices in retail [10].

The overall aim of our project is to investigate the link between different management practices and productivity. This strand of work focuses on simulating various in-store scenarios grounded in empirical case studies with a leading UK retailer. In this paper we aim to understand and predict how Agent-Based Modeling and Simulation (ABMS) can assess and optimize the impact of people management practices on customer satisfaction and Word-Of-Mouth (WOM) in relation to the performance of a service-oriented retail department. To achieve this aim we have adopted a case study approach using applied research methods to collect both qualitative and quantitative data. In summary, we have worked with a leading UK retail organization to conduct four weeks' of informal participant observations in four departments across two retail stores, forty semi-structured interviews with employees including a sixty-three-item questionnaire on the effectiveness of retail management practices, and drawn upon a variety of established information sources internal to the company. This approach has enabled us to acquire a valid and reliable understanding of how the real system operates, revealing insights into the working of the system as well as the behavior of and interactions between the different individuals and their complementary roles within the retail department. Using this knowledge and data, we have applied ABM to devise a functional representation of the case study departments. By executing the developed ABS models we can run experiments to investigate the effects of different management scenarios.

So far we have only studied the impact of people management practices (e.g. training and empowerment) on a customer base that is not influenced by any external or internal stimuli, and hence does not evolve [11, 12]. In order to be able to investigate the impact of management practices on customer satisfaction in a more realistic way we need to consider the cues that stimulate customers to respond to these practices. Therefore our focus is currently on building the capabilities to model customer evolution as a consequence of the implementation of management practices. Consumer decision making, and consequently changes in the behavior of customers over time, are driven by an integration of each consumer's cognitive and affective skills [13]. At any given point in time a customer's behavior, as the product of an individual's cognitions, emotions, and attitudes, may be attributable to an external social cue such as a friend's recommendation or in-store stimuli [14], or an internal cue such as memory of one's own previous shopping experiences [15]. Changing customer requirements may in turn alter what makes a successful management practice (because these are context specific, and customers are a key component of any retail context). In order to enable such studies we enhanced our original Management Practices Simulation (ManPraSim) model and introduced some new features.

In this paper we discuss the key features we have implemented in order to allow the investigation of these kinds of behavioral dynamics. What we are learning here about modeling human behavior has implications for modeling any complex system that involves many human interactions and where the people function with some degree of autonomy.

## 2. Related Work

When investigating the behavior of complex systems the choice of an appropriate modeling technique is very important [16]. To inform the choice of modeling technique, the relevant literature spanning the fields of Economics, Social Science, Psychology, Retail, Marketing, OR, Artificial Intelligence, and Computer Science has been reviewed. Within these fields a wide variety of approaches are used which can be classified into three main categories: analytical methods, heuristic methods, and simulation. Often a combination of these are used within a single model (e.g. [17, 18]. After a thorough investigation of the relevant literature we have identified simulation as being the most appropriate approach for our purposes.

Simulation introduces the possibility of a new way of thinking about social and economic processes, based on ideas about the emergence of complex behavior from relatively simple activities [19]. The purpose of simulation is either to better understand the operation of a target system, or to make predictions about a target system's performance [20]. It allows the testing and evaluation of a theory, and investigation of its implications. Whereas analytical models typically aim to explain correlations between variables measured at one single point in time, simulation models are concerned with the development of a system over time [21]. They allow us to study the transient conditions during the process, along the way to the eventual outcome [22]. The outcome of simulation study is the acquisition of more knowledge about the dynamic behavior of a system, i.e. the state changes of the model as time advances [23]. There are many different approaches to OR simulation, amongst them Discrete-Event Simulation (DES), System Dynamics (SD), and ABS, which is sometimes also referred to as individual-based simulation [24]. The choice of the most suitable approach depends on the issues investigated, the input data available, the required level of analysis, and the type of answers that are sought [16].

Although computer simulation has been used widely since the 1960s, ABMS only became popular in the early 1990s [25]. It is described by [26] as a mindset as much as a technology: 'It is the perfect way to view things and understand them by the behavior of their smallest components'. In agent-based models a complex system is represented by a collection of agents that are programmed to follow simple behavioral rules. Agents can interact with each other and with their environment to produce complex collective behavioral patterns. The main characteristics of agents are their autonomy and their ability to take flexible action in reaction to their environment, both determined by motivations generated from their internal states. They are designed to mimic the attributes and behaviors of their real-world counterparts.

ABS is still a relatively new simulation technique and its principal application has been in academic research. There are, however, a wide range of existing application domains that are making use of the agent paradigm and developing agent-based systems, for example in software technology, robotics, and complex systems, and an ever increasing number of computer games use the ABMS approach. [27] draw a distinction between two central Multi-Agent System (MAS) paradigms: multi-agent decision systems and multi-agent simulation systems. In multi-agent decision systems, agents participating in the system must make joint decisions as a group. Mechanisms for joint decision-making can be based on economic mechanisms, such as an auction, or alternative mechanisms, such as argumentation. Multi-agent simulation systems are used as a model to simulate some real-world domain. The typical application is in domains which involve many different components, interacting in diverse and complex ways and where the system-level properties are not readily inferred from the properties of the components. ABMS is extensively used by the game and film industry to develop realistic simulations of individual characters and societies. It is used in computer games, for example The SIMS™ [28], or in films when diverse heterogeneous characters animations are required, for example the Orcs in Lord of the Rings™ [29].

Due to the characteristics of the agents, the ABMS approach (which is often implemented as a process interaction DES) appears to be more suitable than the event scheduling modeling approach (which is typically associated with the term DES) when it comes to modeling human-centered systems. This is due to the fact that it supports the modeling of autonomous behavior. ABMS seems to promote a natural form of modeling these systems [22]. There is a structural correspondence between the real system and the model representation, which makes them more intuitive and easier to understand than for example a system of differential equations as used in SD. [30] emphasizes that one of the key strengths of ABMS is that the system as a whole is not constrained to exhibit any particular behavior as the system properties emerge from its constituent agent interactions. Consequently, assumptions of linearity, equilibrium and so on, are not needed.

On the other hand, some difficulties have been linked to use of ABMS. One disadvantage that is often mentioned is the high demand in computational power compared to other simulation techniques, most notably in that all of the interactions between agents and between the agent and the environment have to be considered during the simulation run. However, rapid growth in affordable computer power makes this problem less significant [31]. There is consensus in the literature that it is difficult to evaluate agent-based models, because of the heterogeneity of the agents and because the behavior of the system emerges from the interactions between the individual entities. [32] argues that for 'intellective models' (e.g. models that illustrate the relative impact of basic explanatory mechanisms) validation is somewhat less critical and the most important thing is to maintain a balance between keeping a model simple and maintaining veridicality (the extent to which a knowledge structure accurately reflects the information environment it represents). More recently some attempts have been made to develop a common approach to assuring the credibility (verification and validation) of ABS models (see for example [33 - 35].

A two stage approach for empirical validation has been proposed by [36]. The first stage comprises micro-validation of the behavior of the individual agents and the authors recommend this is conducted with reference to real data on individual behavior. The second stage covers macro-validation of the aggregate or emergent behavior of the model when individual agents interact and is advised to be conducted with reference to aggregate time series data. Further, [36] remark that at the macro level only qualitative validation judgments may be possible. Also, problems often occur through the lack of adequate empirical data. [37] state that most quantitative research has concentrated on 'variable and correlation' models that do not cohere well with process-based simulation that is inherent in agent-based models. Finally, [37] point out the danger that people new to ABMS may expect too much from the models, particularly regarding their predictive ability. To mitigate this problem it is important to be clear with individuals about what this modeling technique can really offer, in order to guide realistic expectations.

## 3. Model Design and Data Collection

Before building a simulation model one needs to understand the particular problem domain [38]. In order to gain this understanding we have conducted some case studies. What we have learned during these case studies is reflected in the conceptual models presented in this chapter. Furthermore we explain how we intend to use the data we have gathered during our case studies. Throughout the rest of the paper we will use the term 'actor' to refer to a person in the real system, whereas the term 'agent' will be reserved for their counterparts in the simulation model.

### 3.1 Knowledge Gathering

Case studies were undertaken in four departments across two branches of a leading UK retailer. The case study work involved extensive data collection techniques, spanning: participant observation, semi-structured interviews with team members, management and personnel, completion of survey questionnaires and the analysis of company data and reports (for further information, see [39]). Research findings were consolidated and fed back (via report and presentation) to employees with extensive experience and knowledge of the four departments in order to validate our understanding and conclusions. This approach has enabled us to acquire a valid and reliable understanding of how the real system operates, revealing insights into the working of the system as well as the behavior of and interactions between the different actors within it.

In order to make sure that our results regarding the application of management practices are applicable for a wide variety of departments we have chosen two types of case study departments which are substantially different not only in their way of operating but also their customer type split and staff setup. We collected our data in the Audio & Television (A&TV) and the WomensWear (WW) departments of the two case study branches.

The key differences between these two department types can be summarized as follows. The average customer service time in A&TV is significantly longer, and the average purchase is significantly more expensive than in WW. The likelihood of a customer seeking help in A&TV is also much higher than in WW. Out of customers who have received advice, those in WW have a higher likelihood of making a purchase (indeed customers' questions tend to be very specific to a desired purchase) than in A&TV. Considering customer types, A&TV tends to attract more solution demanders and service seekers, whereas WW customers tend to be shopping enthusiasts. Finally, it is important to note that the conversion rate (the likelihood of customers making a purchase) is higher in WW than in A&TV.

*3.2 Conceptual Modeling*

A conceptual model captures only the essential and relevant characteristics of a system. It is formulated from the initial problem statement, informal user requirements, and data and knowledge gathered from analysis of previous development models [23]. We have used the knowledge gained from the case studies to inform our conceptual models of the system to be investigated, the actors within the system, and their behavioral changes due to certain stimuli.

*3.2.1 Main Concepts for the Simulation Model*

Our initial ideas for the simulation model and its components are display below in Figure 1. Here we can see a bird's eye view of a shopping centre with customers (red dots) visiting a range of shops. Based on this conceptual overview, we determined that we required three types of agents (customers, sales staff and managers), each with a different set of relevant attributes. Global parameters can influence any aspect of the system. The core of our conceptual simulation model consists of an ABS model with a user interface to allow some form of user interaction (change of parameters) before and during runtime. Regarding system outputs, we aim to find some emergent behavior on a macro level. Visual representation of the simulated system and its actors allows us to monitor and better understand the interactions of entities within the system. Coupled with the standard DES performance measures, we can then identify bottlenecks to assist with optimization of the modeled system.

<Insert Figure 1 somewhere here>

*3.2.2 Concepts for the Agents*

We have used state charts for the conceptual design of our agents. State charts show the different states an entity can be in and define the events that cause a transition from one state to another. This is exactly the information we need in order to represent our agents at a later stage within the simulation environment. We have found this form of graphical representation a useful part of the agent design process because it is easier for an expert in the real system (who is not an expert in ABMS) to quickly take on board the model conceptualization and provide useful validation of the model structure and content.

Designing and building a model is to some extent subjective, and the modeler has to selectively simplify and abstract from the real scenario to create a useful model [40]. A model is always a restricted copy of the real system, and an effective model consists of only the most important components of the real system. In our case, the case studies indicated that the key system components take the form of the behaviors of an actor and the triggers that initiate a change from one behavior to another. We have developed state charts for all of the agents in our simulation model. Figure 2 shows as an example the conceptual template model of a customer agent. The transition rules have been omitted to keep the chart comprehensible. They are explained in detail in the Section 3.3.

<Insert Figure 2 somewhere here>

*3.2.3 Concepts for a Novel Performance Measure*

We have introduced a customer satisfaction level index as a novel performance measure using satisfaction weightings. This new measure is required because existing indices such as queuing times or service times are less useful in modeling services than manufacturing activities. In essence, purely quantitative measures fail to capture the quality of service, which is arguably the most important potential trade-off with retail productivity. The inevitable trade-off between quality and quantity is particularly salient when customers and staff come face-to-face and therefore we consider this measure of quality in conjunction with others of quantity.

Historically customer satisfaction has been defined and measured in terms of customer satisfaction with a purchased product [41]. The development of more sophisticated measures has moved on to incorporate customers' evaluations of the overall relationship with the retail organization, and a key part of this is the service interaction. Indeed, empirical evidence suggests that quality is more important for customer satisfaction than price or value-for-money [42], and extensive anecdotal evidence indicates that customer-staff service interactions are an important determinant of quality as perceived by the customer.

The index allows customer service satisfaction to be recorded throughout the simulated lifetime. The idea is that certain situations can have a bigger impact on customer satisfaction than others, and therefore weights can be assigned to events to account for this. Satisfaction indices do not change the likelihood that a customer revisits the department, and although this may at first appear counter-initiative, empirical evidence supports the notion that regardless of a customer's changing shopping intentions he or she will tend to repeat shopping habits [43]. Applied in conjunction with an ABS approach, we expect to observe interactions with individual customer differences; variations which have been empirically linked to differences in customer satisfaction (e.g. [44]). This helps the analyst to find out to what extent customers underwent a positive or negative shopping experience and it also allows the analyst to put emphasis on different operational aspects and try out the impact of different strategies.

*3.2.4 Concepts for Modeling Customer Evolution*

There are two different ways in which we consider customer evolution: external stimulation attributable to the WOM and internal stimulation triggered by memory of one's own previous shopping experiences (this is still work in progress). Sharing information with other individuals (referred to as WOM), significantly affects the performance of retail businesses [45]. An important source of WOM results from customer experiences of retail outlets, and a customer's judgment about whether or not the experience left them feeling satisfied (or dissatisfied). We incorporate WOM in our simulation model by using the number of satisfied customers at the end of the day to calculate the number of additional customers visiting on the next day. The calculation takes into account that only a fraction of people act upon received WOM.

Our concept for representing internal stimulation comprises the exertion of influence on picking certain customer types more often than others. An enthusiastic shopper with a high satisfaction score is much more likely to go shopping more frequently than a disinterested shopper. Therefore, we have introduced some constraints (e.g. out of all customers picked 50% have to be enthusiastic shoppers, 30% normal shoppers, and 20% disinterested shoppers, or, customers with a higher satisfaction score are more likely to be picked to revisit the department). [15] identified different types of internal states and environmental stimuli which act as consumer cues for triggering impulse buying. Internal cues include respondents' positive and negative feeling states, and environmental cues include atmospheric cues in retail settings, marketer-controlled cues, and marketing mix stimuli. When modeling internal stimulation we specifically examine the impact of a consumer's memory of his or her own shopping experience and the impact on department performance measures.

*3.3 Empirical Data*

Often agents are based on analytical models or heuristics and, in the absence of adequate empirical data, theoretical models are employed. However, we use frequency distributions to model state change delays and probability distributions to model decision making processes because statistical distributions are the best way in which we can represent the numerical data we have gathered during our case study work. In this way a population is created with individual differences between agents, mirroring the variability of attitudes and behaviors of their real human counterparts.

The frequency distributions are modeled as triangular distributions defining the time that an event lasts, using the minimum, mode, and maximum duration and these figures are based on our own observations and expert estimates in the absence of objective numerical data.

The probability distributions are partly based on company data (e.g. the rate at which each shopping visit results in a purchase) and partly on informed estimates (e.g. the patience of customers before they leave a queue). Table 1 and 2 show some of the distributions we have defined for our simulation models. We also gathered some company data about work team numbers and work team composition, varying opening hours and peak times, along with other operational details.

**Table 1.** Sample frequency distribution values

| event | probability it occurs |
|---|---|
| someone makes a purchase after browsing | 0.37 |
| someone requires help | 0.38 |
| someone makes a purchase after getting help | 0.56 |

**Table 2.** Sample probability values

| situation | min | mode | max |
|---|---|---|---|
| leave browse state after … | 1 | 7 | 15 |
| leave help state after … | 3 | 15 | 30 |
| leave pay queue (no patience) after … | 5 | 12 | 20 |

**4. Implementation**

*4.1 Implementation of the Main Concepts*

Our ManPraSim model has been implemented in AnyLogic™ (version 5.5) which is a Java™ based multi-paradigm simulation software [46]. During the implementation we have applied the knowledge, experience and data accumulated through our case study work. Within the simulation model we can represent the following actors: customers, service staff (with different levels of expertise) and different types of managers. Figure 3 shows a screenshot of the customer and staff agent logic in AnyLogic™ as it has been implemented in the latest version of our ManPraSim model (v4). Boxes represent states, arrows transitions, arrows with a dot on top entry points, circles with a B inside branches, and numbers satisfaction weights.

<Insert Figure 3 somewhere here>

At the beginning of each simulation run the main customer pool is created which represents a population of potential customers who visit the simulated department on an unspecified number of occasions. Once the simulation has started customers are chosen at a specified rate (customer arrival rate) and released into the simulated department. Currently two different customer types are implemented: customers who want to buy something and customers who require a refund. If a refund is granted, a customer decides whether his or her goal changes to leaving the department straight away, or to making a new purchase. The customer agent template consists of four main blocks which all use a very similar logic. In each block, in the first instance, customers try to obtain service directly from an employee and if they cannot obtain it (i.e. no suitable staff member is available) he or she has to queue. The customer is then either served as soon as the suitable staff member becomes available, or leaves the queue if they do not want to wait any longer (an autonomous decision). A complex queuing system has been implemented to support different queuing rules. Once customers have finished their shopping (either successfully or not) they leave the simulated department and are returned back to the main customer pool where he or she rests until they are picked again.

While the customer is in the department a satisfaction score is calculated by summing the satisfaction weights attached to the transitions that take place during the customer's visit. For example, a customer starts browsing and then requires some help. If he or she obtains help immediately his or her satisfaction score goes up (+2) and after he or she received the help the score goes up again (+4). He or she then moves to the till. If he or she has to wait for help and then leaves the queue because he or she is fed up waiting his or her score goes down (-2). Upon leaving the department he or she will end up in this example with an overall satisfaction score of +4 due to the good and prompt advisory service. The actual customer satisfaction weights shown in Figure 3 were derived based on our experience during the data collection period and validated via interviews with the store management staff.

In comparison to the customer agent state chart, the staff agent state chart is relatively simple. Whenever a customer requests a service and the staff member is available and has the right level of expertise for the task requested, the staff member commences this activity until the customer releases the staff member. Whereas the customer is the active component of the simulation model, the staff member is currently passive, simply reacting to requests from the customer (although both are technically implemented as active objects).

*4.2 Key Features*

There are some additional key features that the simulation model possesses which we describe in the following two sections. First we provide an overview of some important features which have already been implemented in an older version of our simulation model (ManPraSim model v2) but are still relevant to modeling the evolution of customers. This sets the scene for the subsequent description of the new features we have implemented in the latest two versions of our simulation model (ManPraSim model v3 + v4).

*4.2.1 Key Features of the ManPraSim Model v2*

In the ManPraSim model v2 we introduced realistic footfall, customer types, a finite population of customers, and a quick exit of customers when the department is closing. In this paper we only provide an overview of these features because they are described in more detail elsewhere (see [47]).

**Realistic footfall (introduced in v2):** There are certain peak times where a relatively high number of customers are in the department and consequently the demands placed on staff members are greater. In the ManPraSim model v2 we have implemented hourly fluctuations through the addition of realistic footfall (based on some automatically recorded sales transaction data) reflecting different patterns (peaks and troughs) of customer footfall during the day and across different days of the week. In addition we can model the varying opening hours on different days of the week.

**Customer types (introduced in v2):** In real life customers display certain shopping behaviors which can be categorized. Hence we enhance the realism of our agents' behavior in the ManPraSim modelv2 by introducing customer types to create a heterogeneous customer base, thereby allowing us to test customer populations closer to what we would find in reality. Customer types have been introduced based on three identified by the case study organization's own market research (shopping enthusiasts, solution demanders, service seekers) and have been expanded by the addition of two further types (disinterested shoppers, and internet shoppers who are customers that only seek advice but are likely to buy only from the cheapest source i.e. the internet). The three customer types have been identified by the case study organization as the customers who make biggest contribution to their business, in terms of both value and frequency of sales. In order to avoid artificially inflating the amount of sales that we model we have introduced the two additional types which use services but do not tend to make purchases. The definition of each type is based on the customer's likelihood to perform a certain action, classified as either: low, moderate, or high. The definitions can be found in Table 3.

**Table 3.** Definitions for each type of customer

| Customer type | Likelihood to | | | |
|---|---|---|---|---|
| | buy | wait | ask for help | ask for refund |
| Shopping enthusiast | high | moderate | moderate | low |
| Solution demander | high | low | low | low |
| Service seeker | moderate | high | high | low |
| Disinterested shopper | low | low | low | high |
| Internet shopper | low | high | high | low |

**Distribution adaptation (introduced in v2):** In the ManPraSim model v2 we have two algorithms which have been developed to imitate the influence of the customer type attributes mentioned above on customer behavior. They have been implemented as methods that are invoked when defining the state change delays modeled by triangular frequency distributions, and when supporting decision making modeled by probability distributions. Basically the methods define new threshold values for the distributions based on the likelihood values mentioned above. Program 1 shows an example of the pseudo code for the probability distribution threshold correction algorithm. If the customer is a shopping enthusiast and is about to make the decision whether to make a purchase or to leave the department directly (see Figure 3, customer state chart, second branch after leaving the browse state) a corrected threshold value (probability) for this decision is calculated. For this calculation the original threshold of 0.37 (see Table 2) is taken into account and, for a shopping enthusiast where there is a high likelihood to buy (see Table 3), the corrected threshold value is calculated as follows: $0.37+0.37/2 = 0.56$. Consequently the likelihood that a shopping enthusiast precedes to the checkout rather than leaving the department without making any purchase has risen by 18.5%.

**Program 1.** Pseudo code for the probability distribution threshold correction algorithm

```
for (each threshold to be corrected) do
{
    if (OT < 0.5) limit = OT / 2 else (limit = 1 − OT) / 2
    if (likelihood = 0) CT = OT − limit
    if (likelihood = 1) CT = OT
    if (likelihood = 2) CT = OT + limit
}

where:   OT = original threshold
         CT = corrected threshold
         likelihood : 0 = low; 1 = moderate; 2 = high
```

**Finite customer population (introduced in v2):** A key aspect to consider is that the most interesting system outcomes evolve over time and many of the goals of the retail company (e.g. service standards) are planned strategically over the long-term. In the ManPraSim model v2 we have therefore introduced a finite population of customers (main customer pool) where each customer agent is assigned certain characteristics based on the customer types mentioned above. The customer type split in the main customer pool can be defined via an initialization file before the execution of the simulation. The shopping experience of each visit (satisfaction index) is stored in the long term memory of the agent after he or she has left the department. In this way the service a customer experiences can be evaluated over the complete simulated time span.

**Quick exit at closing time (introduced in v2):** In the ManPraSim model v2 we have added transitions that emulate the behavior of customers when the store is closing. These transitions result in an immediate exit of each customer for his or her current state (i.e. the equivalent to a customer running out of shopping time and promptly exiting the store). Not all customer states have these additional transitions because it is for example very unlikely that a customer will leave the store immediately when they are already queuing to pay. Advanced model development means that now the simulated department empties within a ten to fifteen minute period, which conforms to what we have observed in the real system.

*4.2.2 Key Features of the ManPraSim Models v3 + v4*

In the ManPraSim model v3 we have introduced a staff pool with an additional staff type, a new operation mode to support sensitivity analyses and the first implementation of customers' WOM featuring a static main customer pool. Furthermore, we have created some new performance measures that allow us to measure and record the shopping experience of individual visits to the department as well as the daily performance of the department. Finally, in the ManPraSim model v4 we have introduced the second implementation of customers' WOM featuring a dynamic main customer pool.

**Staff pool and additional staff types (introduced in v3):** Retail trends reflect that shops are now open for longer hours over more days of the week, and our case study organization is no exception. To accurately incorporate this source of system variability, we have introduced a staff pool in the ManPraSim model v3 to allow different staffing on different days of the week. The simulation uses Full Timers (FT) to cover all staff shifts required during weekdays. Additional staff who are required to cover busy weekend shifts are modeled by Part Timers (PT). The maximum number of staff required of each type is calculated during the simulation initialization. A staff pool is then created to include weekday FT staff of different types, and the required generic PT to fill the gaps left in the staff shifts which need to be covered. At the beginning of each day the simulation checks how many workers are required and picks the required amount of staff out of the pool at random. The selection process is ordered as follows: FT first and then PT, if required. PT workers have been defined as a generic staff type and can take over any required role. We have tried to model a staff rota with more complex constraints, for example FT staff working five days followed by two days off, however this has as yet proved unsuccessful. Therefore the currently modeled shifts do not incorporate days off work for FT staff.

**Noise reduction mode (introduced in v3):** In the ManPraSim model v3 a noise reduction mode has been implemented which allows us to conduct a sensitivity analysis with constant customer arrival rates, constant staffing throughout the week and constant opening hours. We have used the average values of real world case study data to define the constant values. This has resulted in different values for the different case study department types. As this is a simulation model we cannot take out all the stochasticity, but this way at least we can reduce the system noise to clearly see the impact of the parameter under investigation. When we progress to reintroduce the system noise, the knowledge we have accumulated when experimenting in noise reduction mode helps us to better understand patterns in systems outcomes, and be better able to attribute causation to the introduction of a particular variable (bearing in mind that in the end we are particularly interested in patterns between variables when they are interacting with one another and not in isolation).

**Word of mouth (introduced in v3+v4):** We have developed two different strategies for modeling WOM (which have been implemented in the ManPraSim model v3 (static main customer pool) and in v4 (dynamic main customer pool). Both strategies use the same algorithm to calculate the additional customers attracted to the department through positive WOM and those lost customers attributable to negative WOM. They differ in the way they pick these customers. In the first case we have used a static main customer pool which is defined with the customer split at the beginning of the simulation run. During the simulation run, the population of the main customer pool stays constant and all customers in the pool are available to be randomly picked to go shopping. In the second case we allow the main customer pool size to change; rather than picking customers from the existing pool we create new customers and add them to the main customer pool. When the main customer pool grows or declines the customer arrival rate is also adjusted, i.e. if the pool grows by 10% due to positive WOM so does the customer arrival rate and vice versa for negative growth.

The algorithm to calculate the number of additional new customers and the lost customers is the same for both strategies. We count the number of customers from the previous day who were satisfied and those who were dissatisfied with the service provided during their shopping visit(s). Satisfied customers are more likely to recommend shopping in the department to others whereas dissatisfied customers are more likely to advise others not to visit. The adoption rate (i.e. the success rate of convincing others to commit to either action) depends on the proportion of people who act upon the received WOM (the adoption fraction) and how many contacts a customer has (the contact rate). The equation behind this calculation is shown below in Equation 1.

**Equation 1.** Calculation of adoption rate

$$n_{\text{additional customers}(d)} = (n_{\text{satisfied}(d-1)} - n_{\text{dissatisfied}(d-1)}) * \text{adoption fraction} * \text{contact rate}$$

where: $d$ = current day
$d-1$ = previous day
$n$ = number of ...

At the beginning of each trading day we create a daily customer pool that contains randomly chosen customers from the main customer pool according to the required number of customers defined by the customer arrival rates for the day which might differ throughout the day (e.g. considering peak times).

For the first strategy (static customer pool) we simply pick the core number of customers (defined by the original customer arrival rates) plus an additional number of customers which is calculated through Equation 1. These customers are also selected from the main customer pool at random. If the number is negative we release customers from our daily customer pool back into the main customer pool. If we have to release all of our customers (due to customer dissatisfaction, which may result from bad service due to a large number of customers visiting on the previous day) the simulation terminates. This means that in a sense that the store is closed for good because it cannot recover from this situation in the current implementation.

The second strategy (dynamic customer pool) is slightly more complex. Here we expand or decrease the main customer pool by the number of additional or lost customers (respectively) calculated at the beginning of each trading day. This differs from the static strategy because we do not use the existing customers to model direct WOM influence (i.e. the ones resulting from the service satisfaction of the previous day). This seems to be closer to reality because the WOM in most cases carries positive messages and therefore attracts new visitors rather than motivating existing customers to come more often. Negative WOM dissuades potential customers from visiting the department in the first place. Our daily customer pool is created considering the hourly customer arrival rates (derived from our case study) as well as the actual size of the main customer pool (see Equation 2).

**Equation 2.** Calculation of core customers per day

$$\text{core customers per day} = \frac{\text{dynamic pool size} * \text{known customers per day}}{\text{static pool size}}$$

where:  customers per day = daily customer pool not including WOM additional customers
case study customers per day = derived from hourly customer arrival rate case study data
dynamic pool size = actual main pool size
static pool size = starting main pool size

Once the daily pool size has been established, customers are chosen from the main customer pool according to the requested customer split and customer pool size. Then additional customers are created as new agents. The type of each new customer is determined at random. They are created with a positive satisfaction score of +1 because they have heard only positive things about the department and therefore will already have a positive attitude toward the shop (otherwise they would not have been attracted to it). Through these additional (or lost) agents the daily customer pool size can expand (or contract).

If the negative WOM causes an overall reduction in customers then we choose customers at random from the daily customer pool. Whether or not a particular customer is permanently removed from the main customer pool is determined as follows, depending on the satisfaction score of the chosen customer. If he or she has a positive satisfaction score then the score is neutralized and the customer is returned to the daily customer pool (i.e. he or she can come back for further shopping visits but the shop must start to satisfy the customer's needs for good service to secure further visits from the customer). If he or she has a neutral or negative score he or she is permanently eliminated from the simulation (i.e. he or she will not come back for any more shopping trips). In the second case the main customer pool size will shrink. If the negative WOM results in more agents being deleted than there are customers available in the daily customer pool the simulation will terminate (and the department cannot recover from this in the current implementation).

For both strategies, when it reaches shop closing time all customers from the daily customer pool are released back into the main customer pool where they are available for the next day. For the second strategy (dynamic customer pool) it is very likely that the main pool size will then change due to WOM-triggered additions or losses of customers whereas for the first strategy (static customer pool) the main pool size remains constant.

**New performance measures:** With the introduction of a finite population (represented by our customer pool) we have had to rethink the way in which we collect statistics about the satisfaction of customers. Previously, the life span of a customer has been a single visit to the department. At the end of his or her visit, the individual's satisfaction score (direction and value) has been recorded. Now the life span of a customer lasts the full runtime of the simulation and he or she can be picked several times to visit the department during that period. Our previous performance measures now collect different information: satisfaction scores considering customers' satisfaction history. These measures do not reflect individuals' satisfaction with the current service experience but instead the satisfaction with the overall service experience during the lifetime of the agent. Furthermore, they are biased to some extent in that an indifferent rating quickly shifts into satisfaction or dissatisfaction (arguably this is realistic because most people tend to make a judgment one way or the other).

Whilst the above is still a valuable piece of information we would also like to know how current service is perceived by each customer. For this reason we have introduced a new set of performance measures to record the experience of each customer's individual visit (referred to as Customer Satisfaction Measure – Experienced Per Visit or *CSM-EPV*). These are the same measures as before but on a day-to-day basis they are not anchored by the customer's previous experiences. We also examine the sum of these 'per visit' scores across the lifetime of all customers (referred to as Customer Satisfaction Measure – Accumulated Historical Data or *CSM-AHD*).

Another new measure tracks the satisfaction growth for customers' current and overall service experience. With the incorporation of varying customer arrival rates, opening hours and staffing we have brought in a set of performance measures that capture the impact of these variations on a daily basis. These measures are particularly useful for optimizing departmental performance throughout the week. We also record the satisfaction growth per each individual customer visit. At the end of the simulation run the simulation model produces a frequency distribution which informs us about how satisfied or dissatisfied individual customers have been with the service provided. Furthermore, all forms of customer queue (cashier, normal help, expert help, and refund decision) are now monitored through new performance measures that record how many people have been queuing in a specific queue, and how many of these lost their patience and left the queue prematurely. This measure helps us to understand individual customers' needs because it tells us what individual customers think about the service provided. Finally, we have added some methods for writing all parameters and performance measures into files to support documentation and analysis of the experiments.

*4.3 Model Validation*

Validation ensures that the model meets its intended requirements in terms of the methods employed and the results obtained. In order to test the operation of the ManPraSim model and ascertain face validity we have completed several experiments.

We have completed validation exercises on the micro and macro level, as proposed by [36]. Micro simulation comprises tracing an individual entity through the system by building a trajectory of an individual's behavior throughout the runtime. We have conducted this test for our two different agent types (customer and staff agents) considering all of the different roles these can take. Furthermore we have validated who triggers which process and for what reason. We have then extensively tested the underlying queuing system. On the macro level we have tried to establish simulation model performance outputs that are sound. It has turned out that conducting the experiments with the input data we collected during our case studies did not provide us with a satisfactory match to the performance output data of the real system. We identified the staffing setup used in the simulation models as the root cause of the problems. The data we use here have been derived from real staff rotas. On paper these real rotas suggested that all workers are engaged in exactly the same (one) type of work throughout the day but we know from working with, and observing workers in, the case study organization that in reality each role includes a variety of activities. Staff members in the real organization allocate their time proactively between competing tasks such as customer service, stock replenishment, and processing purchases. Proactive behavior refers to the individual staff members' self-started, long term oriented, and persistent service behavior which goes beyond explicit job demands [48].

So far our simulation models incorporate only one type of work per staff member. For example, the A&TV staff rota indicates that only one dedicated cashier works on weekdays. When we have attempted to model this arrangement, customer queues have become extremely long, and the majority of customers ended up losing their patience and leaving the department prematurely with a high level of dissatisfaction. In the real system we observed other staff members working flexibly to meet the customer demand, and if the queue of customers grew beyond a certain point then one or two would step in and open up further tills to take customers' money before they became dissatisfied with waiting. Furthermore, we observed that a service staff member, when advising a customer, would often continue to close the sale (e.g. filling in guarantee forms and receiving payment from the customer) rather than asking the customer to queue at the till for a cashier whilst moving on to the next customer.

This means that currently our abstraction level is too high and we do not model the real system in an appropriate way. We hope to be able to fix this in a later version. For now we do not consider this to be a big problem so long as we are aware of it. We model as an exercise to gain insights into key variables and their causes and effects and to construct reasonable arguments as to why events can or cannot occur based on the model; we model for insights, not precise numbers.

In our experiments we have modulated the staffing levels to allow us to observe the effects of changing key variables but we have tried to maintain the main characteristic differences between the departments (i.e., we still use more staff in the WW department compared to the A&TV department, only the amount has changed).

## 5. Experiments

The purpose of the experiments described below is to further test the behavior of the simulation models as they become increasingly sophisticated, rather than to investigate management practices per se. We have defined a set of standard settings (including all probabilities, staffing levels and the customer type split) for each department type which we use as a basis for all our experiments. The standard settings exclude the main customer pool sizes for the different departments which are determined in the first experiment and added to the set of standard settings for subsequent experiments. For the experiments that involved only a static main customer pool the run length has been 10 weeks. For the experiments that involved a dynamic main customer pool or both types the run length has been 52 weeks. We have conducted 20 replications (in most cases) to allow the application of rigorous statistical tests.

*5.1 Comparing Customer Pool Sizes*

The first experiment is a sensitivity analysis. We want to find out what impact the customer pool size has on the simulation results and which of our performance measures are affected by it. For this experiment we have used our new noise reduction mode (described in Section 4.2.2). It allows us to focus our attention on the impact of the variable to be investigated. We have run the experiments separately for both case study department types using the standard settings and a pool size ranging from 2,000 to 10,000 customers in increments of 2,000. We will interpret the results in terms of customer-related performance measures, which can be separated out between the standard measures which are defined explicitly in terms of at which stage of their visit a customer leaves the department, and those related to customer satisfaction and dissatisfaction.

**Hypothesis 1:** We predict that varying the customer pool size will not significantly change standard customer measures nor those customer satisfaction measures related measured in terms of experience per visit (CSM-EPV).

**Hypothesis 2:** We predict that the customer satisfaction measures which accumulate historical data (CSM-AHD) will result in significantly more customers being categorised as either 'satisfied' or 'dissatisfied' when compared to the number of customers in these categories for the customer satisfaction measure CSM-EPV.

**Hypothesis 3:** We predict that the customer pool size will significantly impact on all CSM-AHD measures, specifically the greater the customer pool size, the higher the count of neutral customers, and the lower the count of satisfied and dissatisfied customers.

Our results in percentage terms (see Table 4) appear to provide support for all hypotheses. In order to rigorously test our hypotheses, statistical analyses were applied to the raw customer counts as follows. To investigate Hypothesis 1, a series of one-way between groups ANOVAs were conducted separately for A&TV and WW. Levene's test for homogeneity of variances was violated by CSM-EPV for satisfied customers in A&TV only, therefore a more stringent significance value was set for this variable ($p<.01$). Given that 8 simultaneous comparisons were to be drawn, a Bonferroni adjustment was made to the alpha level, tightening the significance levels to $p<.00625$ (from $p<.05$) and $p<.00125$ (from $p<.01$). For A&TV, no significant differences were found in the standard performance measures or CSM-EPV between different customer pool sizes. For WW, no significant differences were found apart from the number of customer leaving before receiving normal help (p=.0059). Inspecting mean differences revealed that the mean customer count with a customer pool of 8,000 was higher than for 4,000, 6,000, and 10,000 customers. However, upon examination of supporting data, no single paired comparison was significantly different. Across customer pools, the mean customer count varied between just 4.6 and 7.5 customers leaving before receiving normal help, and so we can explain this pattern in terms of the relative infrequency of this event and conclude that the difference is of little practical significance. This leads us to conclude that overall the evidence supports Hypothesis 1. We predicted this pattern because customer decisions are driven by customer types and the proportional mix of each type is kept constant throughout the experiment.

To test Hypothesis 2, a series of independent samples T-tests were conducted to evaluate the differences between the two methods of measuring customer satisfaction. For A&TV, a significantly higher count of customers were categorized as satisfied on CSM-AHD (M=18,031.18, SD=824.11) as compared to their counterparts on the CSM-EPV measure (M=15,071.12, SD=108.47), t(102.43)=35.61, p<.000 (equal variances not assumed). The same pattern was reflected in a significantly higher count of dissatisfied customers (M=17,134.13, SD=895.06) on CSM-AHD as compared to CSM-EPV (M=11,404.65, SD=221.34), t(111.06)=62.14, p<.000 (equal variances not assumed). The effect sizes are very large in both cases, yielding .86 and .95 respectively.

For WW, a significantly higher count of customers were categorized as satisfied on CSM-AHD (M=54,161.66, SD=3,559.94) as compared to on CSM-EPV (M=33,508.33, SD=82.76), t(99.11)=58.00, p<.000 (equal variances not assumed). The inverse occurred for the count of dissatisfied customers, whereby a smaller number were counted on CSM-AHD (M=3,447.49, SD=1,049.66) than CSM-EPV (M=4,626.00, SD=128.59), t(101.97)=-11.14, p<.000 (equal variances not assumed). The effect size was very large for A&TV at .94, and medium-to-large at .39 for WW.

Overall as expected CSM-AHD generally results in a significantly higher count of satisfied and dissatisfied customers than CSM-EPV. This phenomenon has been discussed in Section 4.2.2. The evidence for A&TV unequivocally supports this hypothesis, whereas it is mixed for WW. The surprisingly relatively low count of dissatisfied of CSM-AHD in WW recurs in evaluation of the next hypothesis (see below).

**Table 4.** Descriptive statistics for Experiment 1 (all to 2 d.p.)

| A&TV | | | | | | | | | | |
|---|---|---|---|---|---|---|---|---|---|---|
| Customer pool size | 2,000 | | 4,000 | | 6,000 | | 8,000 | | 10,000 | |
| Customers … | Mean | SD | Mean | SD | Mean | SD | Mean | SD | Mean | SD |
| … leaving after making a purchase | 29.5% | 0.1% | 29.5% | 0.1% | 29.6% | 0.1% | 29.5% | 0.1% | 29.4% | 0.1% |
| … leaving before receiving normal help | 2.7% | 0.2% | 2.7% | 0.2% | 2.8% | 0.2% | 2.7% | 0.2% | 2.8% | 0.2% |
| … leaving before receiving expert help | 1.1% | 0.1% | 1.1% | 0.1% | 1.1% | 0.0% | 1.1% | 0.0% | 1.1% | 0.0% |
| … leaving whilst waiting to pay | 17.5% | 0.3% | 17.4% | 0.3% | 17.3% | 0.4% | 17.4% | 0.3% | 17.4% | 0.3% |
| … leaving before finding anything | 49.3% | 0.4% | 49.3% | 0.4% | 49.3% | 0.3% | 49.3% | 0.3% | 49.3% | 0.3% |
| … leaving satisfied (accumulated historical data) | 46.5% | 1.9% | 45.3% | 1.2% | 43.8% | 0.8% | 42.9% | 0.6% | 42.0% | 0.9% |
| … leaving neutral (accumulated historical data) | 8.7% | 0.2% | 12.1% | 0.3% | 14.6% | 0.2% | 16.5% | 0.2% | 18.1% | 0.3% |
| … leaving dissatisfied (accumulated historical data) | 44.8% | 2.2% | 42.5% | 1.4% | 41.6% | 1.1% | 40.6% | 1.0% | 39.9% | 1.1% |
| … leaving satisfied (experience per visit) | 36.8% | 0.3% | 36.9% | 0.2% | 36.9% | 0.2% | 36.8% | 0.2% | 36.7% | 0.4% |
| … leaving neutral (experience per visit) | 35.3% | 0.3% | 35.3% | 0.3% | 35.3% | 0.3% | 35.2% | 0.3% | 35.2% | 0.2% |
| … leaving dissatisfied (experience per visit) | 27.9% | 0.6% | 27.8% | 0.5% | 27.8% | 0.6% | 27.9% | 0.5% | 28.0% | 0.6% |
| WW | | | | | | | | | | |
| Customer pool size | 2,000 | | 4,000 | | 6,000 | | 8,000 | | 10,000 | |
| Customers … | Mean | SD | Mean | SD | Mean | SD | Mean | SD | Mean | SD |
| … leaving after making a purchase | 46.5% | 0.1% | 46.5% | 0.1% | 46.5% | 0.1% | 46.6% | 0.1% | 46.5% | 0.1% |
| … leaving before receiving normal help | 0.0% | 0.0% | 0.0% | 0.0% | 0.0% | 0.0% | 0.0% | 0.0% | 0.0% | 0.0% |
| … leaving before receiving expert help | 0.1% | 0.0% | 0.1% | 0.0% | 0.1% | 0.0% | 0.1% | 0.0% | 0.1% | 0.0% |
| … leaving whilst waiting to pay | 10.0% | 0.2% | 10.0% | 0.3% | 10.0% | 0.3% | 10.0% | 0.2% | 10.0% | 0.2% |
| … leaving before finding anything | 43.4% | 0.3% | 43.4% | 0.3% | 43.3% | 0.3% | 43.3% | 0.3% | 43.3% | 0.2% |
| … leaving satisfied (accumulated historical data) | 93.4% | 0.3% | 88.3% | 0.3% | 84.0% | 0.3% | 80.7% | 0.3% | 78.0% | 0.3% |
| … leaving neutral (accumulated historical data) | 3.9% | 0.1% | 7.2% | 0.2% | 10.2% | 0.2% | 12.6% | 0.2% | 14.7% | 0.2% |
| … leaving dissatisfied (accumulated historical data) | 2.7% | 0.2% | 4.5% | 0.3% | 5.8% | 0.2% | 6.7% | 0.3% | 7.3% | 0.3% |
| … leaving satisfied (experience per visit) | 52.4% | 0.2% | 52.5% | 0.1% | 52.5% | 0.1% | 52.5% | 0.1% | 52.5% | 0.1% |
| … leaving neutral (experience per visit) | 40.3% | 0.3% | 40.3% | 0.2% | 40.2% | 0.3% | 40.2% | 0.2% | 40.2% | 0.2% |
| … leaving dissatisfied (experience per visit) | 7.3% | 0.2% | 7.2% | 0.3% | 7.2% | 0.2% | 7.2% | 0.2% | 7.3% | 0.2% |

To investigate Hypothesis 3, a series of one-way between groups ANOVAs were conducted. Levene's test for homogeneity of variances was violated by CSM-AHD for satisfied and dissatisfied customers in A&TV and only dissatisfied customers in WW, therefore a more stringent significance value was set for these variables ($p<.01$). Further to this a Bonferroni adjustment was implemented because of the multiple comparisons, tightening the significance value to $p<.0167$ (from $p<.05$) and $p<.0033$ (from $p<.01$).

Separately for A&TV and WW, all 3 ANOVAs for each CSM-AHD variable exhibited significant differences ($p<.000$). Examining mean differences for A&TV: for satisfied customers, significant differences were present between all paired pool sizes except for between 2,000 and 4,000; 6,000 and 8,000; and 8,000 and 10,000. The mean customer count uniformly decreased as the customer pool size increased, with significant differences focused on comparisons between paired customer pool sizes with a greater size differential. For neutral customers, all paired comparisons were significantly different ($p<.000$), and examination of mean differences reveals a uniformly increasing trend in the customer count as the customer pool size increases. For dissatisfied customers, significant differences persisted (at $p<.0033$) between the following pairs: 2,000 and all other pool sizes; 4,000 and 8,000; and 4,000 and 10,000. As customer pool size increased, the count of dissatisfied customers uniformly decreased.

Looking at mean differences for WW: for satisfied customers, all paired comparisons were significantly different ($p<.000$), and an examination of mean differences revealed a uniformly decreasing trend in the count of satisfied customers as customer pool size increases. For neutral customers, again all paired comparisons exhibited significant differences ($p<.000$), and this time the mean customer counts revealed a uniformly increasing trend as customer pool size increases. Finally, for dissatisfied customers, all paired comparisons were significantly different ($p<.000$), and the mean customer counts behaved according to a uniformly increasing trend as customer pool size increases. These results follow a similar pattern to that presented in A&TV, apart from the increasing number of dissatisfied customers. We expect that this finding can be explained in terms of an accumulation of the impact of a number of subtle differences in the ability of the department to cope with greatly increased numbers of customers (which one-by-one did not produce significant effects, see evaluation of Hypothesis 1).

Looking at the percentages presented in Table 4, we can see that the bigger the customer pool size the more the CSM-AHD values tend to the CSM-EPV values. This can be explained by the fact that with a larger population the likelihood that a specific customer enters the department repeatedly (and therefore accumulates some historical data) reduces. If we sufficiently increase the customer population we expect both customer satisfaction measures to show similar results because the majority of customers are only picked once during the simulation runtime and therefore most customers will enter the department with a neutral satisfaction score rather than with an accumulated score. In summary for Hypothesis 3 we can conclude that the evidence largely supports our prediction, with the exception of dissatisfied customer counts in WW for which we have presented an explanation.

The results for Experiment 1 demonstrate that it is therefore important to select and maintain one customer pool size to ensure that all performance measures are providing comparable information for different experiments. Our case study organization does not collect or hold data on customer pool size so we have instead calculated a suitable value based on the average numbers of customers who visit the department per day (585 for A&TV and 915 for WW) and an estimate of customers' average inter-arrival time (two weeks for A&TV and one week for WW). These values have been estimated considering the standard customer type split for the corresponding department as well as customer demand for the items sold in that department. They do not necessarily apply to other customer type splits. Using these values we have calculated an appropriate customer pool size for each department (8,000 for A&TV and 6,500 for WW). We use these customer pool sizes for all subsequent experiments.

*5.2 Comparing Normal and Noise Reduction Mode*

In our second experiment we want to investigate the importance of considering hourly differences in customer arrival rates and daily differences in staffing and opening hours (normal mode). These features have been added to investigate hypotheses related to the real performance of the case study departments and specific patterns that occur on a day-to-day basis. Modeling this level of detail can be problematic when we conduct a sensitivity analysis where we want to be able to attribute causation to the introduction of a particular variable. For this kind of experiment we prefer to control some of the system noise to clearly see the impact of the parameter under investigation (noise reduction mode).

**Hypothesis 4:** We predict that the standard customer performance measures will not significantly differ across the two modes.

**Hypothesis 5:** We predict the two modes will produce significantly different values when looking at the daily performance measures on a day-to-day basis (rather than using averages).

<Insert Figure 4 somewhere here>

**Table 5.** Descriptive statistics for Experiment 2 (all to 2 d.p.)

|  | Adoption fraction Customers … | AF = 0 | | AF = 0.5 | | AF = 1 | | T-Test | | |
|---|---|---|---|---|---|---|---|---|---|---|
|  |  | Mean | SD | Mean | SD | Mean | SD | t-value | p-value | Eta$^2$ |
| A&TV | Overall customer count [per day] | 584.17 | 6.04 | 601.09 | 8.27 | 609.38 | 29.09 | -7.10 | 0.00 | 0.04 |
|  | … leaving after making a purchase [runtime] | 12,062.95 | 36.97 | 12,094.30 | 46.75 | 12,063.05 | 50.29 | -0.01 | 0.99 | 0.00 |
|  | … leaving before receiving normal help [runtime] | 1,133.80 | 67.36 | 1,343.30 | 63.92 | 1,706.55 | 99.01 | -21.39 | 0.00 | 0.14 |
|  | … leaving before receiving expert help [runtime] | 464.65 | 18.25 | 457.00 | 18.47 | 463.55 | 22.27 | 0.17 | 0.87 | 0.00 |
|  | … leaving whilst waiting to pay [runtime] | 7,048.70 | 105.59 | 7,520.30 | 116.47 | 7,633.60 | 124.13 | -16.05 | 0.00 | 0.02 |
|  | … leaving before finding anything [runtime] | 20,182.00 | 134.56 | 20,661.65 | 145.08 | 20,789.75 | 176.33 | -12.25 | 0.00 | 0.00 |
| WW | Overall customer count [per day] | 910.58 | 6.25 | 1,093.24 | 22.44 | 1,224.79 | 39.98 | -65.00 | 0.00 | 0.66 |
|  | … leaving after making a purchase | 29,696.75 | 44.48 | 30,097.55 | 31.34 | 30,256.45 | 45.29 | -39.43 | 0.00 | 0.03 |
|  | … leaving before receiving normal help | 5.40 | 3.30 | 29.55 | 7.24 | 84.60 | 17.71 | -19.66 | 0.00 | 0.81 |
|  | … leaving before receiving expert help | 83.60 | 9.04 | 130.75 | 12.48 | 170.40 | 14.53 | -22.68 | 0.00 | 0.67 |
|  | … leaving whilst waiting to pay | 6,291.95 | 165.73 | 13,148.10 | 200.16 | 18,050.20 | 164.96 | -224.88 | 0.00 | 0.68 |
|  | … leaving before finding anything | 27,662.80 | 166.79 | 33,120.80 | 198.49 | 37,173.90 | 196.79 | -164.89 | 0.00 | 0.30 |

Looking at Figure 4 it is clear that there is very little variation between the two different operation modes across the range of standard performance measures. Examining the descriptives (see Table 5), in most cases the runtime performance measures in both modes are approximately the same with a small number of exceptions. Contrary to hypothesis 4, in WW approximately 25% more customers leave whilst waiting to pay in the normal mode (which consequently influences both customer satisfaction measures), as opposed to the noise reduction mode. This apparent cashier bottleneck appears to be exacerbated by any combination of the three factors which are held constant in noise reduction mode. Further analysis is required to isolate the precise cause. We can also see that, as predicted, the smaller the values the more they differ (on an absolute basis) between the two modes as they are accumulated from a smaller number of events and therefore the influence of different random number streams is more apparent.

<Insert Figure 5 somewhere here>

Examining the daily measures on a day-to-day basis for A&TV in Figure 5, we can observe clear differentiation between the number of customers and transactions across weekdays, Saturdays and Sundays in normal mode whereas the noise reduction mode, as expected, shows no clear patterns. This additional information can be very useful for optimizing the system. For example, we have the lowest number of transactions on Sundays (day 1, 8, etc.) although we do not have the lowest number of customers on this day, therefore we must have a problem with the optimizing the staffing arrangement on Sundays, because on average more customers leave the shop on Sundays without buying anything, despite the fact that the probability of this remains the same for all days of the week.

Overall the experiment has shown that it is legitimate to use the noise reduction mode within the scope of a sensitivity analysis to test the impact of a specific factor on system behavior. It is only when we are interested in specific features that are not available in noise reduction mode (e.g. when we want to study differences between particular days of the week or when we need to obtain quantitative data to optimize a real system) that we need to choose the normal mode.

*5.3 Evaluation the Implementation of WOM*

In the first part of this experiment we want to investigate how customer satisfaction can be influenced by using different ways of representing inter-linkages within a customer population. We compare increasingly sophisticated forms of implementing WOM. Firstly we have used a static customer pool where customer shopping behavior is random (i.e. not influenced by customer's word of mouth). Secondly, we have introduced a static customer pool where shopping behavior is influenced by word of mouth. Finally, we employ a dynamic customer pool to allow us to investigate growth (or reduction) in the size of the overall customer pool (or customer base).

We have chosen a runtime of 52 weeks because we expect to see model developments further into the model runtime, in particular for the dynamic WOM implementation. For this experiment we have used our new noise reduction mode (described in Section 4.2.2) and the following WOM settings: an adoption fraction of 0.5 and contact rate of 2. We expect there to be a link between customers' shopping behavior and customer satisfaction scores, moderated by the type of retail department (i.e. an interaction effect). Specifically, we predict that:

**Hypothesis 6:** In the dynamic conditions, customer pool size will increase during the simulation run for each department up to a maximum value, and maintain this level for the run duration.

**Hypothesis 7:** Average daily customer numbers will be significantly different for each of the 6 WOM conditions. In particular, we predict that WW will experience a higher number of customer visits than A&TV, the more advanced the WOM implementation the higher the number of customer visits, and we expect to observe a positive moderating effect whereby an interaction between the two can explain further increases in average daily customer numbers.

**Hypothesis 8:** Within each department the average number of satisfied customers will significantly decrease, whereas the average number of dissatisfied customers will significantly increase (both CSM-EPV), with the sophistication of the WOM implementation.

The results are presented in multiple forms to allow detailed analysis. Figures 6 to 9 show the results for a single simulation run on a day to day basis, which are integral to revealing day-to-day fluctuations. Due to concerns about the quality of presentation, this will be restricted to the first 26 weeks of each simulation run. The authors have examined graphs for the full runtime period and variables' trends continued as would be expected based on the first 26 weeks. Table 6 shows the descriptive statistics of the average results of the experiment; tabulated values are daily averages across all replications.

Regarding Hypothesis 6, Figure 6 clearly displays that under dynamic conditions the customer pool size grows over the early stages of the simulation until it reaches a maximum value and then stabilizes around this level (for averages see Table 6). Interestingly, the graph displays that WW reaches a notably higher value than A&TV, which we can understand in terms of the tighter staffing constraints in A&TV restricting the availability of staff to serve customers (and hence constrain the growth of the customer population). In WW, the growth in customer numbers can be alikened to the diffusion of customers' word-of-mouth in the real world; as the numbers grow, the rate of growth also increases. In contrast, the growth in customer numbers triggered by WOM in A&TV has a steadily slowing rate of growth, which may be related to the department's limited capacity to satisfy growing customer numbers.

**Table 6.** Descriptive statistics for Experiment 3 (all to 2 d.p.)

| A&TV | | | | | | |
|---|---|---|---|---|---|---|
| Customer pool type | static pool, no WOM | | static pool, WOM | | dynamic pool, WOM | |
| Customers … | Mean | SD | Mean | SD | Mean | SD |
| … leaving after making a purchase | 62,679.65 | 88.11 | 63,224.00 | 69.60 | 63,711.70 | 98.50 |
| … leaving before receiving normal help | 4,727.70 | 105.62 | 10,760.25 | 134.67 | 25,787.85 | 273.77 |
| … leaving before receiving expert help | 2,245.15 | 34.55 | 2,635.05 | 58.72 | 2,969.45 | 55.50 |
| … leaving whilst waiting to pay | 37,010.95 | 195.20 | 47,887.25 | 270.52 | 57,511.75 | 426.80 |
| … leaving before finding anything | 105,929.80 | 311.08 | 119,330.35 | 268.35 | 144,318.30 | 629.55 |
| … leaving satisfied (accumulated historical data) | 165,413.55 | 657.24 | 153,231.20 | 722.13 | 124,970.40 | 971.28 |
| … leaving neutral (accumulated historical data) | 13,501.05 | 209.10 | 17,211.90 | 240.75 | 24,612.60 | 267.97 |
| … leaving dissatisfied (accumulated historical data) | 33,678.65 | 740.80 | 73,393.80 | 803.15 | 144,716.05 | 851.11 |
| … leaving satisfied (experience per visit) | 81,322.90 | 219.71 | 76,344.00 | 171.84 | 69,782.75 | 130.40 |
| … leaving neutral (experience per visit) | 100,754.70 | 430.03 | 122,440.55 | 333.06 | 156,304.80 | 640.76 |
| … leaving dissatisfied (experience per visit) | 30,516.20 | 205.27 | 45,052.35 | 232.58 | 68,211.50 | 213.43 |
| Overall no. of customers | 212,593.25 | 447.40 | 243,836.90 | 286.48 | 294,299.05 | 750.04 |
| Average no. of visits per customer | 26.80 | 5.16 | 30.50 | 5.28 | 3.03 | 0.70 |
| WW | | | | | | |
| Customer pool type | static pool, no WOM | | static pool, WOM | | dynamic pool, WOM | |
| Customers … | Mean | SD | Mean | SD | Mean | SD |
| … leaving after making a purchase | 152,433.10 | 131.28 | 156,100.00 | 83.91 | 159,585.35 | 91.58 |
| … leaving before receiving normal help | 0.40 | 0.60 | 2.40 | 1.93 | 17,590.80 | 600.72 |
| … leaving before receiving expert help | 112.15 | 11.87 | 217.80 | 13.26 | 1,870.45 | 35.82 |
| … leaving whilst waiting to pay | 26,591.25 | 400.65 | 83,259.55 | 361.07 | 199,493.30 | 1,027.04 |
| … leaving before finding anything | 152,807.05 | 432.68 | 204,547.10 | 601.35 | 373,335.45 | 1,121.25 |
| … leaving satisfied (accumulated historical data) | 318,549.55 | 629.39 | 409,429.25 | 896.42 | 382,430.25 | 2,072.47 |
| … leaving neutral (accumulated historical data) | 8,495.55 | 159.34 | 12,185.85 | 144.87 | 41,828.45 | 571.67 |
| … leaving dissatisfied (accumulated historical data) | 4,898.85 | 275.81 | 22,511.75 | 686.50 | 327,616.65 | 1,494.42 |
| … leaving satisfied (experience per visit) | 164,114.20 | 151.98 | 180,283.15 | 183.70 | 175,356.25 | 404.53 |
| … leaving neutral (experience per visit) | 146,188.90 | 435.49 | 196,015.65 | 594.29 | 405,963.70 | 1,179.03 |
| … leaving dissatisfied (experience per visit) | 21,640.85 | 342.18 | 67,828.05 | 277.59 | 170,555.40 | 309.05 |
| Overall no. of customers | 331,943.95 | 601.55 | 444,126.85 | 616.05 | 751,875.35 | 1,296.40 |
| Average no. of visits per customer | 51.10 | 7.23 | 68.50 | 7.27 | 5.87 | 0.48 |

A two-way between groups ANOVA was conducted to test Hypothesis 7. Levene's test for homogeneity of variances was violated, and so we introduced a more stringent significance level (p<.01). The ANOVA revealed a statistically significant main effect for department type [$F(1,114)=3,698,165$, p=.000] and WOM implementation [$F(2,114)=1,226,034$, p=.000]. The effect sizes were extremely high in both cases (partial eta-squared = 1.00). Examining mean differences by department type, WW's customer count was significantly higher than for A&TV. Tukey's post-hoc comparisons were applied to the WOM implementation variable, and all paired comparisons were significant. Results indicated that the more sophisticated the WOM implementation, the higher the customer count. There was also a significant interaction effect between the two independent variables [$F(2,114)=572858.8$, p=.000] with a very high effect size (partial eta-squared = 1.00). This finding suggests that department type moderates the effect of the WOM implementation, resulting in increasingly higher customer counts for WW (as compared to A&TV) with increases in the sophistication of WOM implementation. These results provide support for Hypothesis 7.

Differences in the daily count of customers visiting the department are presented in Figure 7. We expected these relationships because WW has an inherently greater capacity to serve an increased number of customers using the same amount of resources due to the relatively low service requirement of the average WW customer. Although A&TV has some scope to increase its customer base, customer requirements for advice and attention constraint the department's ability to deal with an increased volume of customers. In terms of different WOM implementations, the more sophisticated the implementation, the more opportunities that customers have to spread the word and encourage other customers to visit. Consequently we expected that more sophisticated versions would continue to grow customer numbers, and the interaction with department type builds on the arguments already put forward.

It appears that the restricted availability of staff to serve customers is preventing a commensurate increase in sales. Key bottlenecks occur in A&TV with the provision of normal help and cashier availability to take payment, and the bottlenecks in WW occur with the same variables and also the provision of expert help (see Table 6). All of these bottlenecks result in a significant rise in customers who leave without finding anything to buy in either department.

<Insert Figure 6 somewhere here>

<Insert Figure 7 somewhere here>

Daily differences in the daily count of CSM-EPV satisfied and dissatisfied customers visiting the department are presented in Figures 8 and 9. A series of one-way between-groups ANOVAs were used to test Hypothesis 8 and evaluate the impact of different WOM implementations on average measures of CSM-EPV. A Bernoulli correction was applied to tighten the significance value to $p<.025$. For A&TV, tests revealed a statistically significant difference between the three WOM implementations for the number of satisfied customers [$F(2, 59)=21,201.95$, $p<.000$] and the number of dissatisfied customers [$F(2, 59)=152,952.60$, $p<.000$]. The effect sizes, calculated using eta-squared, were both very high at 1.00. Tukey's post-hoc comparisons identified significant differences between every single paired comparison in the directions predicted ($p<.000$).

For WW, Levene's test for homogeneity of variances was violated for satisfied customers, and so the p-value for this variable was further restricted to $p<.005$. The ANOVA indicated a statistically significant difference between the three WOM implementations for the number of satisfied customers [$F(2, 59)=18,689.89$, $p<.000$] and the number of dissatisfied customers [$F(2, 59)=1,203,536$, $p<.000$]. The effect sizes were very high (eta-squared 1.00 for both). Tukey's post hoc tests revealed significant differences between all paired comparisons ($p<.000$). All were in the predicted direction, with the exception of a large increase in the number of satisfied customers between the first and second implementations of WOM. This exception may be attributable to some slack in the WW staffing being taken up as the level of WOM sophistication (and hence potential customer pool) increased. However, with further WOM developments (dynamic customer pool) the staffing constraints start to take effect and the number of satisfied customers decreases. Looking at Figure 8, it is clear that on a day-to-day basis there are great fluctuations in customer satisfaction. In Figure 9, again much day-to-day variation can be seen, and the significant increase in the average number of dissatisfied customers in the dynamic WW model can be clearly observed.

In general, Hypothesis 8 has been supported, in that evidence suggests that the average number of satisfied customers (CSM-EPV) will decrease with a more sophisticated WOM implementation, and that the number of dissatisfied customers will increase (CSM-EPV). The exceptional behavior of the count of satisfied customers in WW has been explained in terms of slack in the staffing of the department, and with further increases to customer number we would expect the corresponding decrease in the number of satisfied customers to continue.

<Insert Figure 8 somewhere here>

<Insert Figure 9 somewhere here>

In the second part of the experiment we want to explore daily fluctuations in customer numbers. We want to find out if there is a set threshold value for WOM, beyond which we would on the next day create an empty department (and hence cause its permanent closure). This could be triggered by the great variation in daily fluctuations, whereby a very high amount of satisfied customers causes an overwhelming volume of customers on the subsequent day. This could result in crucially high levels of dissatisfaction and result in no customers visiting the department on the day after.

For this experiment we have used our new noise reduction mode (described in Section 4.2.2). We vary the adoption fraction using a step interval of 0.2 whilst keeping the contact rate constant at 5. The high contact rate value has been selected to reach the upper boundary of customer numbers in a reasonable amount of trials.

**Hypothesis 9:** We predict that the higher the WOM adoption fraction, the greater the increase and corresponding subsequent decrease in the number of customer visits. We expect this relationship to be linked to a reduction in satisfaction per customer. We hypothesize that this pattern will vary over time because the impact of WOM will vary on a daily basis.

**Hypothesis 10:** We predict that there is a threshold upon which the daily number of customers shrinks to 0 and consequently the department ceases trading. We expect that this threshold will occur at an earlier stage for A&TV than WW.

Given that we are looking at daily differences, it is important to examine what is going on graphically. Therefore hypotheses will be tested through evaluation of simulation output in terms of daily fluctuations. Comparing average values dampens this variability which we are interested in (due to model stochasticity), and so we will be assess hypotheses using figures and not statistical tests.

Figure 10 presents the customer count relating to different adoption fractions across WW and A&TV (static model). Comparing the two departments, a greater increase in customer numbers can be seen between experimental conditions in WW than in A&TV as the customer pool grows, which is to be expected. This is because in addition to the customer service limitations of A&TV, the starting customer pool size in WW is 23% greater than that of A&TV.

Assessing Hypothesis 9, Figures 10 (and also Figures 11 and 12) suggests that the higher the adoption fraction the greater the day-to-day fluctuations in customer numbers. Each individual line tends to vary a roughly equal amount above and below its middle, adding weight to the hypothesis that each increase in customer numbers is accompanied by a subsequent decrease. Further evidence supporting this hypothesis is covered in assessment of Hypothesis 10.

Evaluating Hypothesis 10, we can see in Figures 10, 11 and 12 that the results for A&TV suggest that it is not able to meet the increased customer demands which the highest adoption fraction (0.6) places on it. The overall shapes of the extreme cases (adoption fraction 0.6) behave in a pronounced cyclic manner for both static and dynamic WOM implementations, with relatively large customer peaks and troughs compared to the other adoption fractions in A&TV. Due to a day with no customers, the A&TV department ceases business. This does appear to be caused by a sharp increase in one day's customers triggering a sharp decrease in the following day's customer numbers. We predicted this relationship because the department's staffing resources remain the same regardless of customer pool growth, and therefore when the daily customer pool grows, the same amount of staffing resources are shared between a larger group of customers. The lower adoption fractions (0.2 and 0.4) result in a relatively restricted variation in customer numbers after approximately 6 weeks in the static implementation, and 15 weeks in the dynamic implementation. It is interesting to compare adoption fractions of 0.4 and 0.6 for the static model because the 0.6 model only attracts a few more customers to trigger its downfall, whereas 0.4 recovers from this near-miss.

Looking at WW in Figures 10 and 11, there are great variations in customer numbers in particular for the highest adoption fraction (0.6) where we can see areas of pronounced oscillating peaks and troughs alternating with less pronounced areas of daily fluctuations. For the dynamic model (Figure 12) at 50 weeks this adoption fraction results in no customers. Given the large amount of variation in customer numbers throughout this simulation, it is not surprising that this phenomenon occurs. Also as observed for A&TV, all adoption fractions produce phases of relatively intense variation in customer numbers alternate with relatively stable phases.

The figures presented support Hypothesis 10 in principle. A&TV the static WOM implementation resulted in zero customers at week 3, whereas in WW the dynamic WOM implementation resulted in zero customers at week 50. This provides some supports the prediction that this would happen at an earlier stage in A&TV than in WW. Further experimentation (not documented here) investigated for each department types we could not finish the simulation run for an adoption fraction value of 0.6 (dynamic pool). This work identified that the boundary adoption fraction to allow the experiment to finish (with the complete runtime of 52 weeks) occurred for the static implementation with a value of up to 0.500 for A&TV, and 0.791 for WW. The higher threshold for WW concords with what we have seen in daily differences. For the dynamic implementation, the adoption fraction can have a value of up to 0.485 for A&TV and up to 0.599 for WW. Any adoption fraction values beyond these result in a premature termination of the department, due to no customers.

In summary, we can conclude that comparing and contrasting different implementations of WOM can provide fruitful insights into department performance. The current models help us to better understand system outcomes where there is a finite staffing resource in a particular department, and the initial customer demands are already ensuring there is little slack in the system (NB we observed this to a greater extent with A&TV than for WW). Investigating a different scenario such as when a retailer has the investment to introduce further resource with increasing customer demand would require a different configuration.

It is clear that the current implementation of a dynamic customer pool results in WOM only directly affecting the growth or decline of the customer pool population on the following day. This means that the behavior of customer count variable results in an unbelievable variation in day-to-day customer figures. In the real world we would expect word of mouth to impact on an individual's behavior with a short delay, perhaps of a couple of days, or until the next opportunity that individual has to visit the department. Not everyone runs to the shop immediately when he or she has received a recommendation, (NB we do not consider that some customers can only shop on certain days of the week, or at certain times of the day), and even if this did happen it is unlikely that a customer passing on WOM would recommend the department to everyone immediately. There is a real delay in the spread of the WOM. Our findings so far suggest that it would be better modeled using a positively skewed distribution showing a large immediate effect and also an ongoing subtler effect which would be strong the more that WOM has accumulated. We will test this in the next version of our simulation model.

<Insert Figure 10 somewhere here>

<Insert Figure 11 somewhere here>

<Insert Figure 12 somewhere here>

## 6. Conclusion

We have presented the conceptual design, implementation and operation of simulation models to help inform the way we which we can understand the impact of people management practices on retail performance. As far as we are aware this is the first time researchers have tried to use an agent-based approach to simulate people management practices such as training and empowerment. Although our simulation model uses specific case studies as source of information, we believe that the general model could be adapted to other retail companies and areas of management practices that have a lot of human interaction.

To respond to our key research questions, we have modeled and simulated customer satisfaction and the effects of WOM diffusion on customer satisfaction. We have compared and analyzed data resulting from two contrasting methods of modeling customer satisfaction: CSM-AHD and CSM-EPV. Likewise, we have adopted a two-pronged approach to customers' WOM and present comparative analysis of different strategies to implement the WOM (using a static and a dynamic main customer pool). The evaluation of daily trends has proved particularly valuable in understanding what causes a department to run out of customers, and further model development is planned to implement a different approach to this.

In this paper we have focused in particular on the capabilities required to model customer evolution as a consequence of the implementation of people management practices. We have discussed conceptual ideas about how to consider external and internal stimuli and have presented an implementation of WOM as one form of external stimuli. We are still testing and calibrating the new features we have implemented. The validation experiments so far have shown that we need to improve our simulation model in order to be able to model the real system in an appropriate way. In particular our current abstraction level with regards to how staff spend their time is much too high. If we want to use real staffing data we need to model how staff allocate their tasks between competing activities rather than focusing on one type of work (i.e. make our staff members proactive). We have already modeled one form of proactive staff behavior (empowerment) but that was only for specific experiments, e.g. (1) a normal sales staff member decides by himself or herself whether to stay or not to stay with a customer when additional expert advice is given which allows them to learn on the job if they wish to do so, or (2) an expert staff member helps out in the provision of normal advice if no customer requires expert help (see [11]). Further development such as self-initiated till manning needs to be implemented in order to mimic the behavior of the staff members in the real system in a more appropriate way.

In addition to continuing to validate our current simulation model we have also planned to experiment with different strategies of modeling customer evolution. We plan to implement and test our conceptual ideas regarding customers' memory of their own shopping experiences (internal stimuli). Furthermore, we have planned to enhance the flexibility of staff members (i.e. empower them) to allow them to respond to customer demand. This will help to solve the staffing problem we have discussed above. In the long term we want to develop our simulation model to support testing the impact of team work related management practices. This looks like an interesting but challenging task because we first need to come up with a way to represent the effects of team work. Furthermore, we would like to enhance the capabilities of our agents, giving them skills in reasoning, negotiation, and co-operation.

In conclusion, we can say that the new features appear promising and we are convinced they have already improved our insights into the operation of the departments within a department store. In particular the new performance measures we collect on a daily basis will be very useful in future for balancing the provision of customer services throughout the week.

Overall, we believe that researchers should become more involved in this multi-disciplinary kind of work to gain new insights into the behavior of organizations. In our view, the main benefit from adopting this approach is the improved understanding of and debate about a problem domain. The very nature of the methods involved forces researchers to be explicit about the rules underlying behavior and to think in new ways about them. As a result, we have brought work psychology and agent-based modeling closer together to form a new and exciting research area.

## 7. Acknowledgements

This work is part of an IDEAS Factory Network Project funded by the UK government through the Engineering and Physical Sciences Research Council (EPSRC) under grant EP/D503949/1. We would also like to thank the reviewers for their fruitful comments on earlier drafts of this article.

## 8. References

[1] Department of Trade and Industry. 2003. *UK productivity and competitiveness indicators*. DTI Economics Paper No. 6. London, UK: DTI. http://www.berr.gov.uk/files/file14767.pdf

[2] Reynolds, J., E. Howard, D. Dragun, B. Rosewell, and P. Ormerod. 2005. Assessing the productivity of the UK retail sector. *International Review of Retail, Distribution and Consumer Research* 15(3):237-280.

[3] Delbridge, R., P. Edwards, J. Forth, P. Miskell, and J. Payne. 2006. *The organisation of productivity: Re-thinking skills and work organisation*. London, UK: Advanced Institute of Management Research. http://www.aimresearch.org/publications/orgprod.pdf.

[4] Siebers, P.O. et al. 2008. *The role of management practices in closing the productivity gap*. AIM Working Paper Series 065. London, UK: Advanced Institute of Management Research. http://papers.ssrn.com/sol3/papers.cfm?abstract_id=1309605.

[5] Wall, T.D. and S.J. Wood. 2005. Romance of human resource management and business performance and the case for big science. *Human Relations* 58(5):429-462.

[6] Pourdehnad, J., K. Maani, and H. Sedehi. 2002. System dynamics and intelligent agent-based simulation: Where is the synergy? In *Proceedings of the 20th International Conference of the System Dynamics Society*, Palermo, Italy.

[7] Birdi, K. et al. 2008. The impact of human resource and operational management practices on company productivity: A longitudinal study. *Personnel Psychology* (in press).

[8] Patel, S. and A. Schlijper. 2004. Models of consumer behaviour. *49th European Study Group with Industry (ESGI 2004)*, Oxford, UK.

[9] Nicholson, M., I. Clarke, and M. Blakemore. 2002. One brand, three ways to shop: Situational variables and multichannel consumer behaviour. *International Review of Retail, Distribution and Consumer Research* 12:131-148.

[10] Keh, H.T., S. Chu, and J. Xu. 2006. Efficiency, effectiveness and productivity of marketing in services. *European Journal of Operational Research* 170(1):265-276.

[11] Siebers, P.O., U. Aickelin, H. Celia, and C. Clegg. 2007a. A multi-agent simulation of retail management practices. In *Proceedings of the Summer Computer Simulation Conference (SCSC 2007)*, San Diego, USA, pp 959-966.

[12] Siebers, P.O., U. Aickelin, H. Celia, and C. Clegg. 2007b. Using intelligent agents to understand management practices and retail productivity. In *Proceedings of the Winter Simulation Conference (WSC 2007)*, Washington DC, USA, pp 2212-2220.

[13] Hansen, T. 2005. Perspectives on consumer decision making: An integrated approach. *Journal of Consumer Behaviour* 4(6):420-437.

[14] Tai, S.H.C. and A.M.C. Fung. 1997. Application of an environmental psychology model to in-store buying behaviour. *International Review of Retail, Distribution and Consumer Research* 7(4):311-337.

[15] Youn, S. and R.J. Faber. 2000. Impulse buying: Its relation to personality traits and cues. *Advances in Consumer Research* 27(1):179-185.

[16] Robinson, S. 2004. *Simulation: The practice of model development and use*. Chichester, UK: John Wiley & Sons

[17] Greasley, A. 2005. Using DEA and simulation in guiding operating units to improved performance. *Journal of the Operational Research Society* 56(6):727-731.

[18] Schwaiger, A. and B. Stahmer. 2003. SimMarket: Multi-agent based customer simulation and decision support for category management. In *Applying Agents for Engineering of Industrial Automation Systems*, edited by M. Shillo et al., LNAI 2831. Berlin: Springer.

[19] Simon, H.A. 1996. *The sciences of the artificial (3rd Ed.)*. Cambridge, MA: MIT Press.

[20] Gilbert, N. and K.G. Troitzsch. 1999. *Simulation for the social scientist*. Open University Press, UK: Buckinghamshire.

[21] Law, A.M., and W.D. Kelton. 1991. *Simulation modeling and analysis (2nd Ed.)*. New York, NY: McGraw-Hill

[22] North, M.J. and C.M. Macal. 2007. *Managing business complexity: discovering strategic solutions with agent-based modeling and simulation*. New York, NY: Oxford University Press.

[23] Garrido, J.M. 2001. *Object-oriented discrete-event simulation with java: A practical introduction*. New York, NY: Kluwer Academic Publishers

[24] Borshchev, A. and A. Filippov. 2004. From system dynamics and discrete event to practical agent based modeling: reasons, techniques, tools. In *Proceedings of the 22nd International Conference of the System Dynamics Society*, Oxford, UK.

[25] Epstein, J.M. and R. Axtell. 1996. *Growing artificial societies: Social science from the bottom up*. Cambridge, MA: MIT Press.

[26] Jeffrey, R. 2008. *Expert voice: Icosystem's Eric Bonabeau on agent-based modeling*. http://www.cioinsight.com/article2/0,3959,1124316,00.asp.

[27] Luck, M., P. McBurney, O. Shehory, and S. Willmott. 2005. *Agent technology: Computing as interaction (a roadmap for agent based computing)*. Liverpool, UK: AgentLink.

[28] ZDNet. 2000. *ZDNet Complexity Digest 2000.10: The Sims and agent based modeling*. http://www.comdig.com/index.php?id_issue=2000.10.

[29] BBC. 2008. *BBC h2g2: Computer animation*. http://www.bbc.co.uk/dna/h2g2/A342 1045/.

[30] Hood, L. 1998. Agent-based modeling. In *Conference Proceedings: Greenhouse Beyond Kyoto, Issues, Opportunities and Challenges*, Canberra, Australia.

[31] Macal, C.M. and M.J. North. 2007. Agent-based modeling and simulation: Desktop ABMS, In *Proceedings of the Winter Simulation Conference (WSC 2007)*, Washington, USA, pp95-106.

[32] Carley, K.M. 1996. *Validating computational models*. CASOS Working Paper. Pittsburgh, PA: Carnegie Mellon University.

[33] Fagiolo, G., P. Windrum, and A. Moneta. 2006. *Empirical validation of agent-based models: A critical survey*. LEM Working Paper 2006/14. Pisa, Italy: Sant'Anna School of Advanced Studies.

[34] Leombruni, R., M. Richiardi, N.J. Saam, and M. Sonnessa. 2006. A common protocol for agent-based social simulation. *Journal of Artificial Societies and Social Simulation* 9(1):15.

[35] Midgley D.F., R.E. Marks, and D. Kunchamwar. 2007. The building and assurance of agent-based models: An example and challenge to the field. *Journal of Business Research* 60(8):884-893.

[36] Moss, S. and B. Edmonds. 2005. Sociology and simulation: Statistical and quantitative cross-validation. *American Journal of Sociology* 110(4):1095-1131.

[37] Twomey, P. and R. Cadman. 2002. *Agent-based modelling of customer behaviour in the telecoms and media markets*. info 4:56-63.

[38] Chick, S.E. 2006. Six ways to improve a simulation analysis. *Journal of Simulation* 1:21-28.

[39] Celia, H. 2007. *Retail management practices and performance: On the shop floor*. Master Thesis. University of the West of England.

[40] Shannon, R.E. 1975. *Systems simulation: The art and science*. Englewood Cliffs, NJ: Prentice-Hall.

[41] Yi, Y. 1990. A critical review of consumer satisfaction, In *Review of Marketing*, edited by V.A. Zeithaml. Chicago, IL: American Marketing Association.

[42] Fornell, C., M.D. Johnson, E.W. Anderson, J. Cha, and B.E. Bryant. 1996. The American customer satisfaction index: Nature, purpose, and findings. *Journal of Marketing* 60(4):7-18.

[43] Ji, M.F. and W. Wood. 2007. Purchase and consumption habits: Not necessarily what you intend. *Journal of Consumer Psychology* 17(4):261-276.

[44] Simon, F. and J.C. Usunier. 2007. Cognitive, demographic and situational determinants of service customer preference for personnel-in-contact over self-service technology'. *International Journal of Research in Marketing* 24(2):163-173.

[45] Marsden, P., A. Samson, and N. Upton. 2005. *The economics of buzz - word of mouth drives business growth finds LSE study*. http://www.lse.ac.uk/collections/pressAndInformationOffice/newsAndEvents/archives/2005/Word_of Mouth.htm.

[46] XJ Technologies. 2008. *Official AnyLogic website*. http://www.xjtek.com/.

[47] Siebers, P.O., U. Aickelin, H. Celia, and C. Clegg. 2007c. Understanding retail productivity by simulating management practices. In *Proceedings of the EUROSIM Congress on Modelling and Simulation (EUROSIM 2007)*, Ljubljana, Slovenia.

[48] Rank, J., J.M. Carsten, J. Unger, and P.E. Spector. 2007. Proactive customer service performance - relationships with individual, task, and leadership variables. *Journal of Human Performance* 20(4):363-390

**Peer-Olaf Siebers** is a research fellow in computer science at The University of Nottingham (UK), where he works in the Intelligent Modelling & Analysis Group.

**Uwe Aickelin** is a full Professor at The University of Nottingham (UK), where he leads the Intelligent Modelling & Analysis Group.

**Helen Celia** is a visiting researcher at the Centre for Organisational Strategy, Learning & Change at Leeds University Business School (UK).

**Chris W. Clegg** is Professor of Organisational Psychology and Director of the Centre for Socio-Technical Systems Design at Leeds University Business School (UK).

**Figures:**

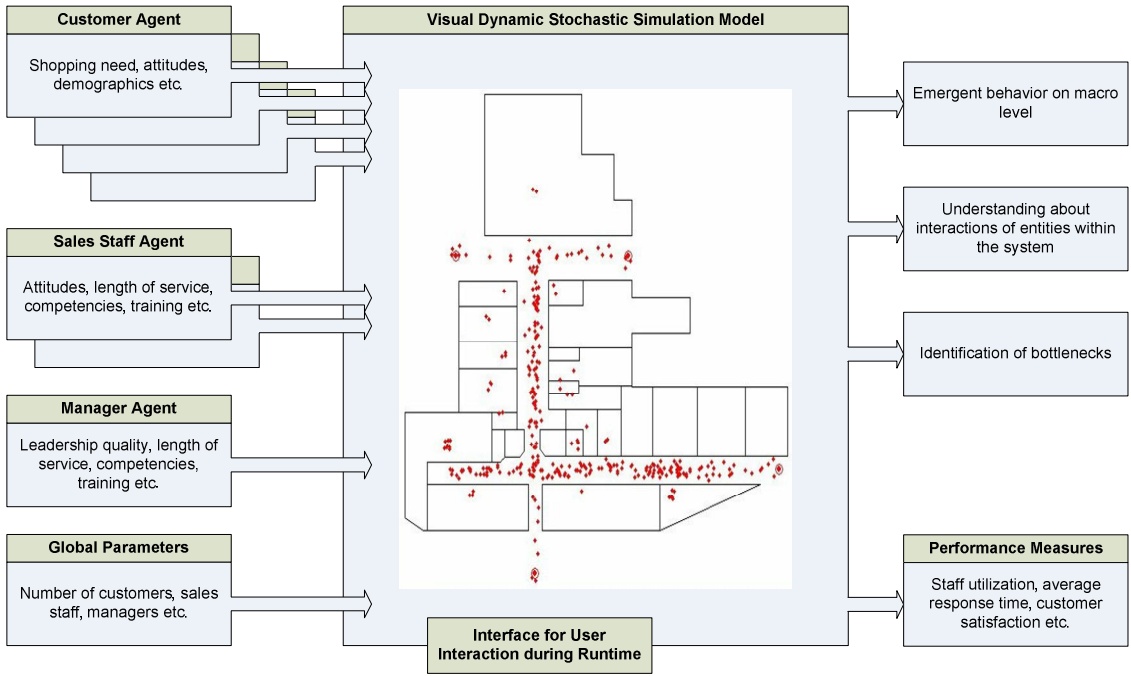

**Figure 1.** Initial ideas for the simulation model and its components

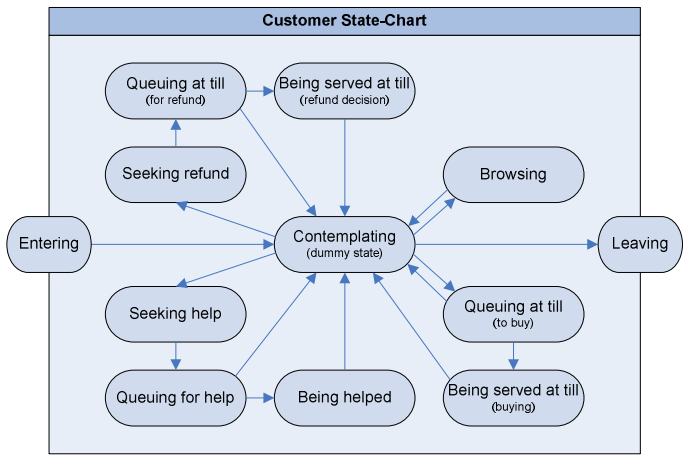

**Figure 2.** Conceptual model of our customer agents (transition rules have been omitted for clarity)

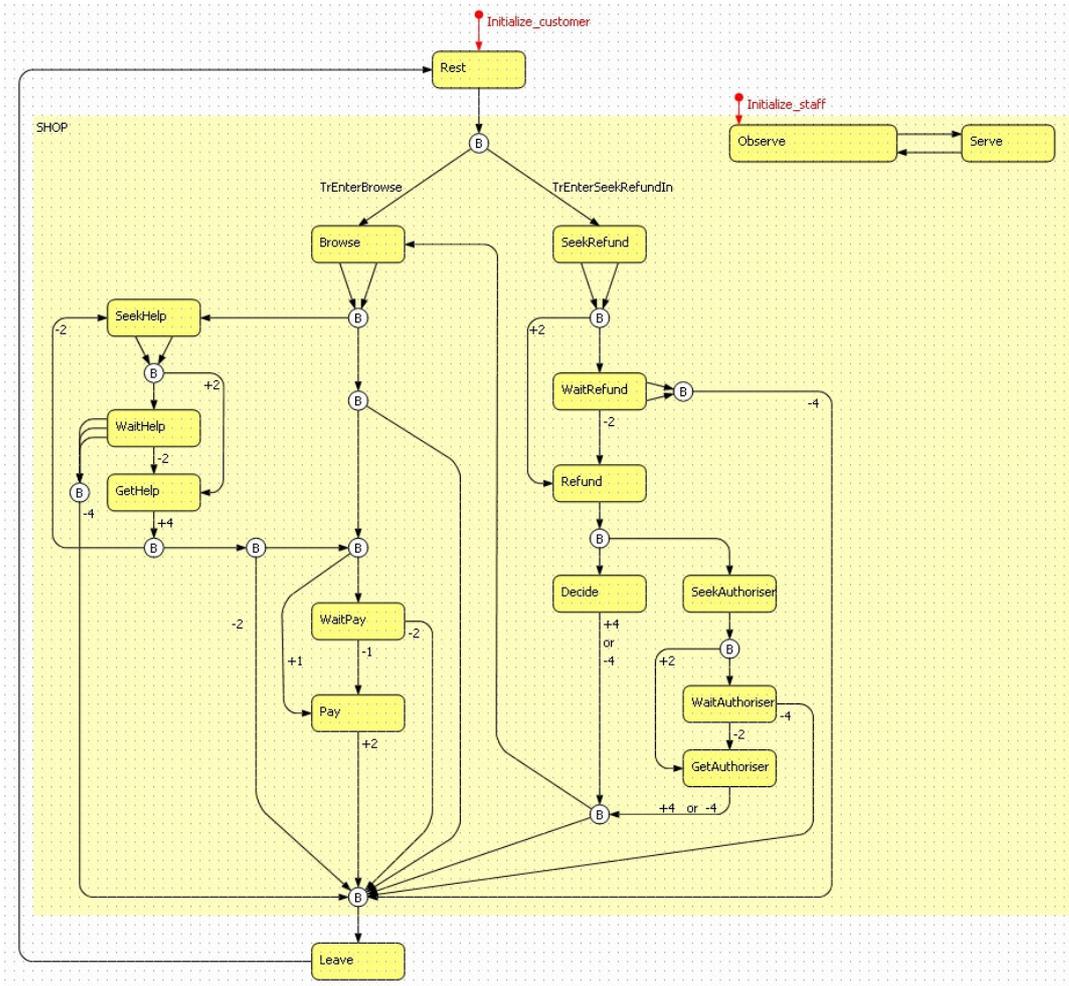

**Figure 3.** Customer (left) and staff (right) agent logic implementation in AnyLogic™

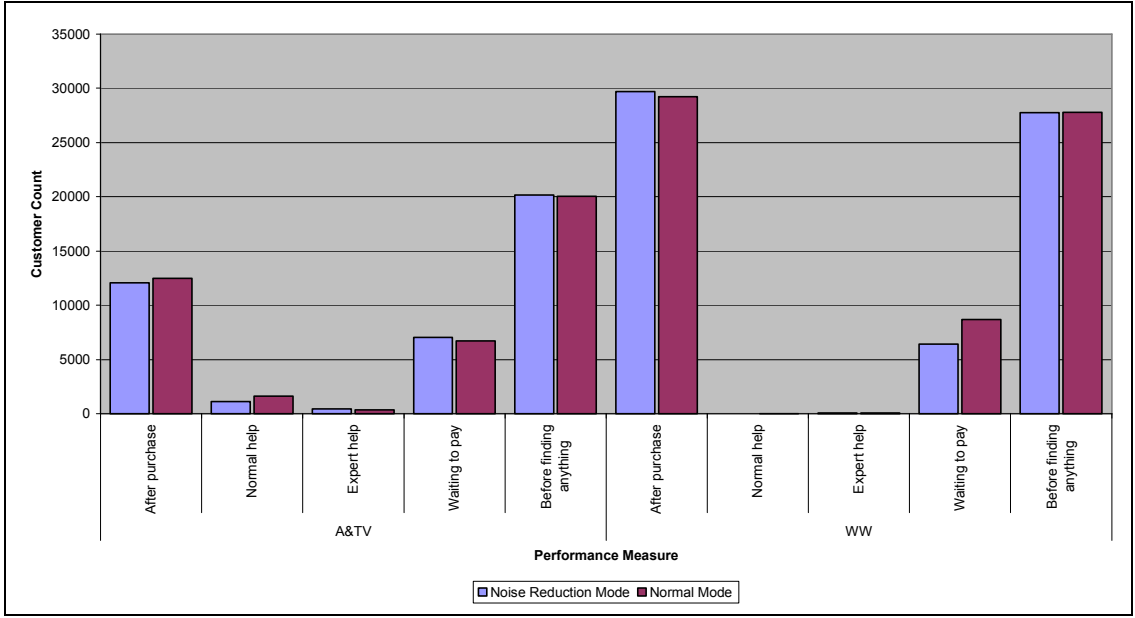

**Figure 4.** Average performance measures, by department and operation mode.

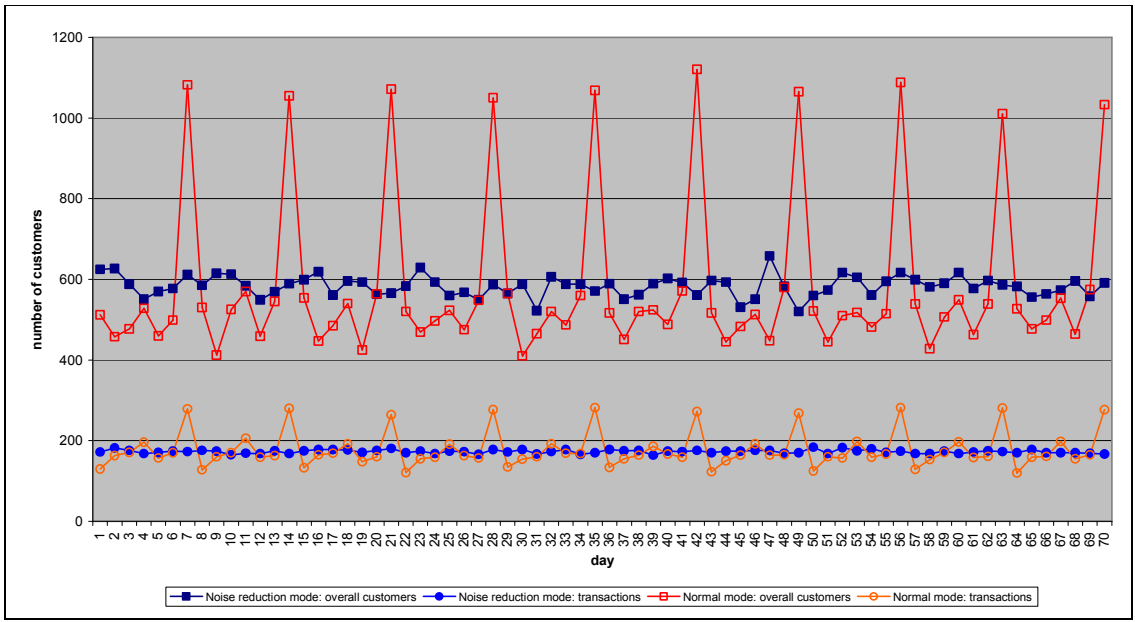

**Figure 5.** Daily customer count and number of transactions in A&TV.

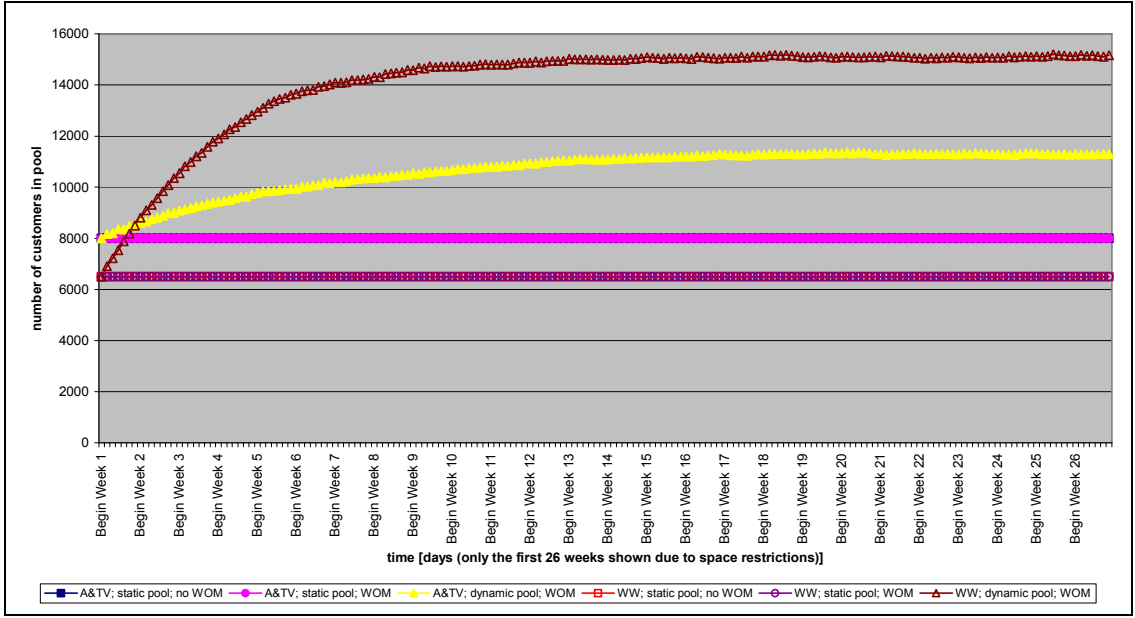

**Figure 6.** Customer pool growth over time.

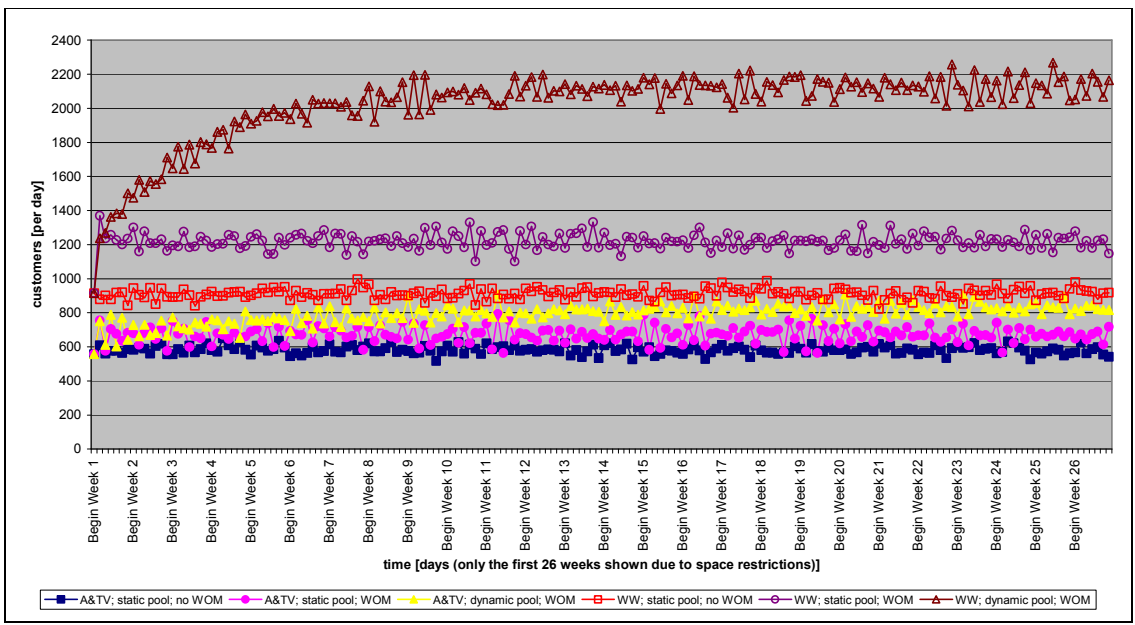

**Figure 7.** Daily count of customers entering the department.

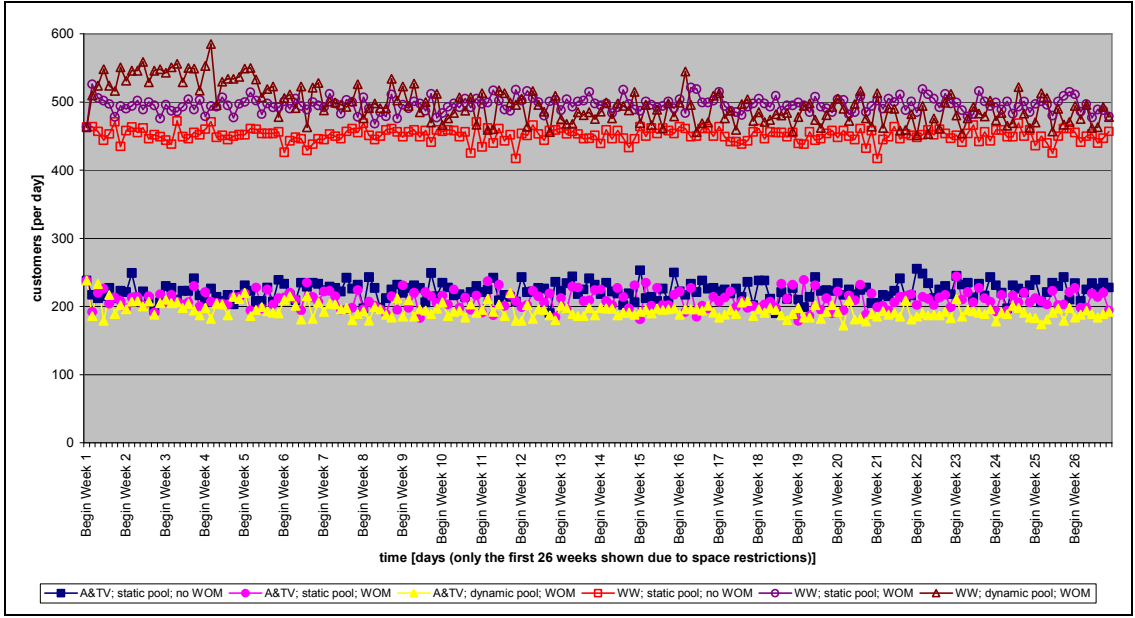

**Figure 8.** Daily count of satisfied customers CSM-EPV.

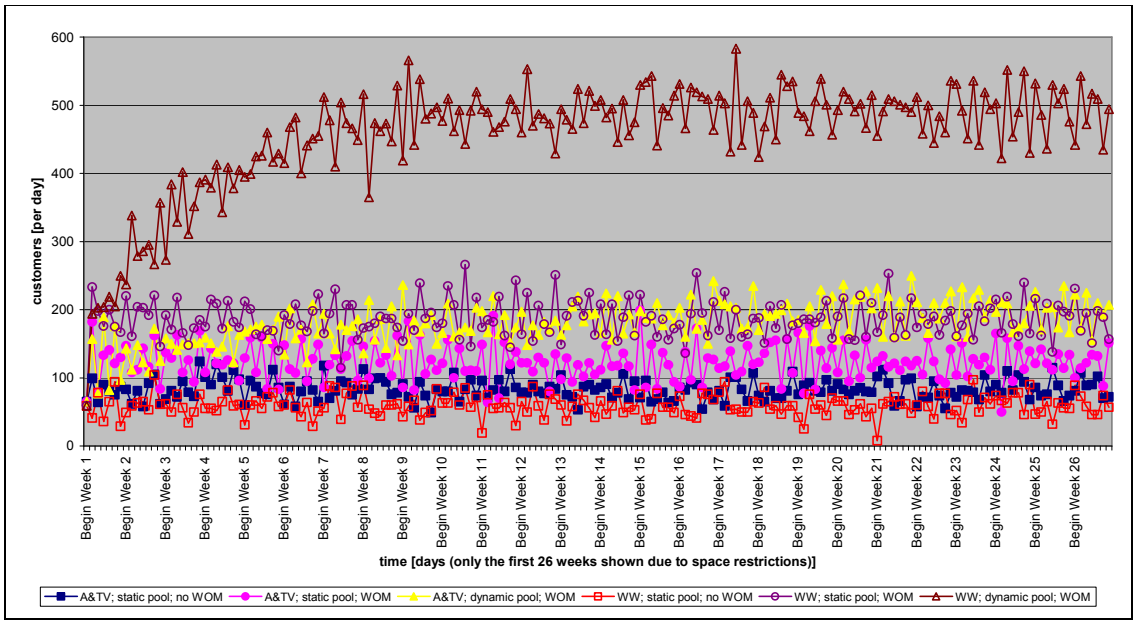

**Figure 9.** Daily count of dissatisfied customers CSM-EPV.

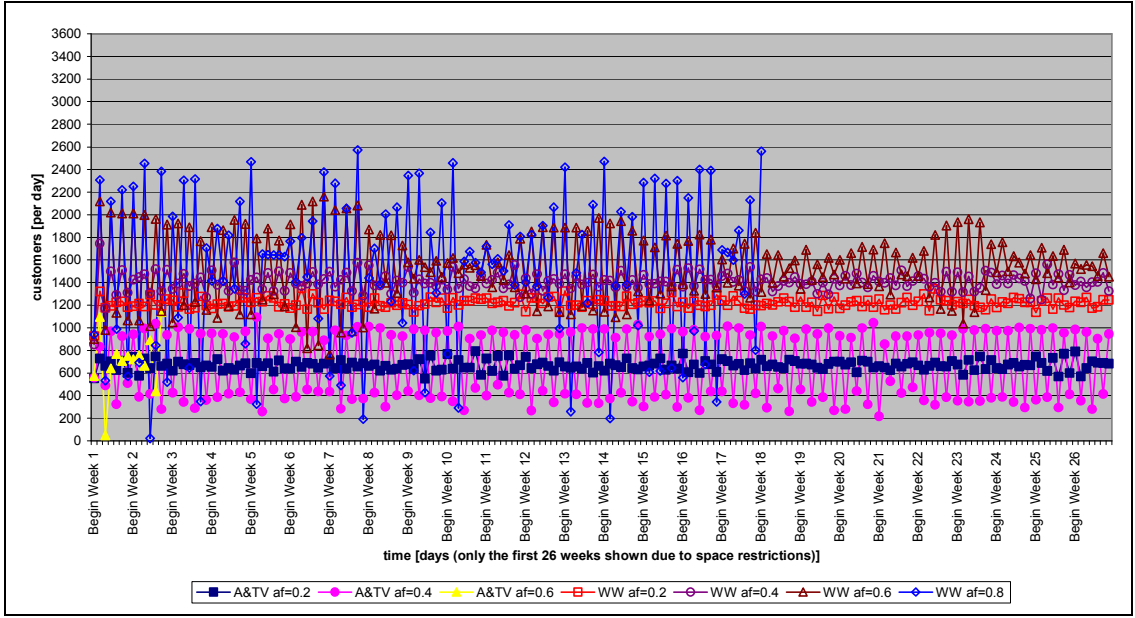

**Figure 10.** Static pool - daily customer count.

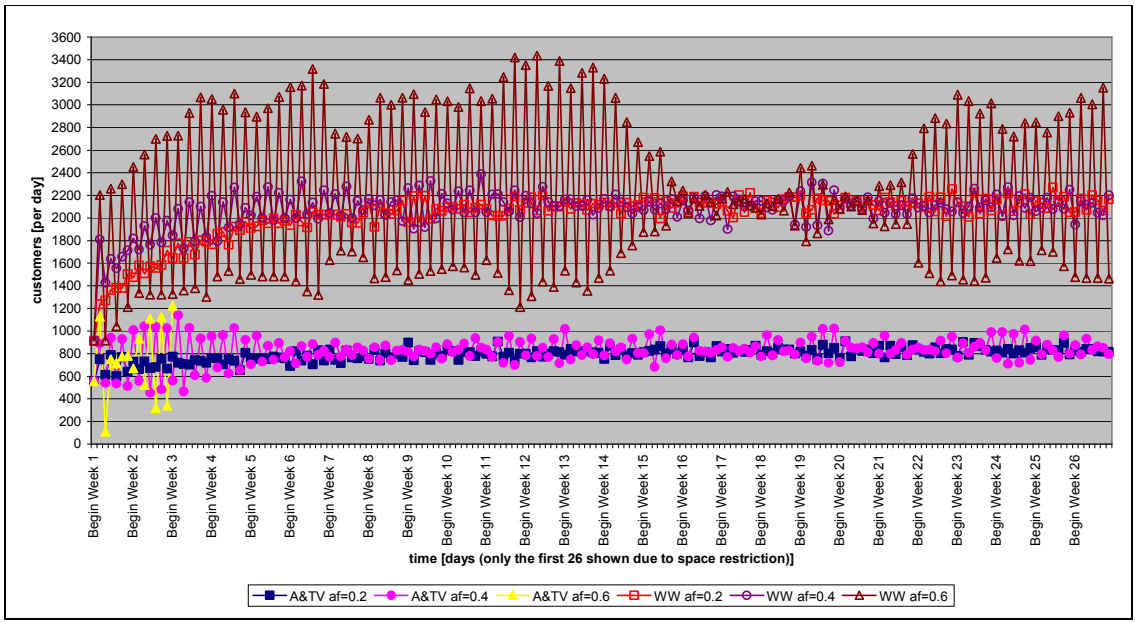

**Figure 11.** Dynamic pool - daily customer count.

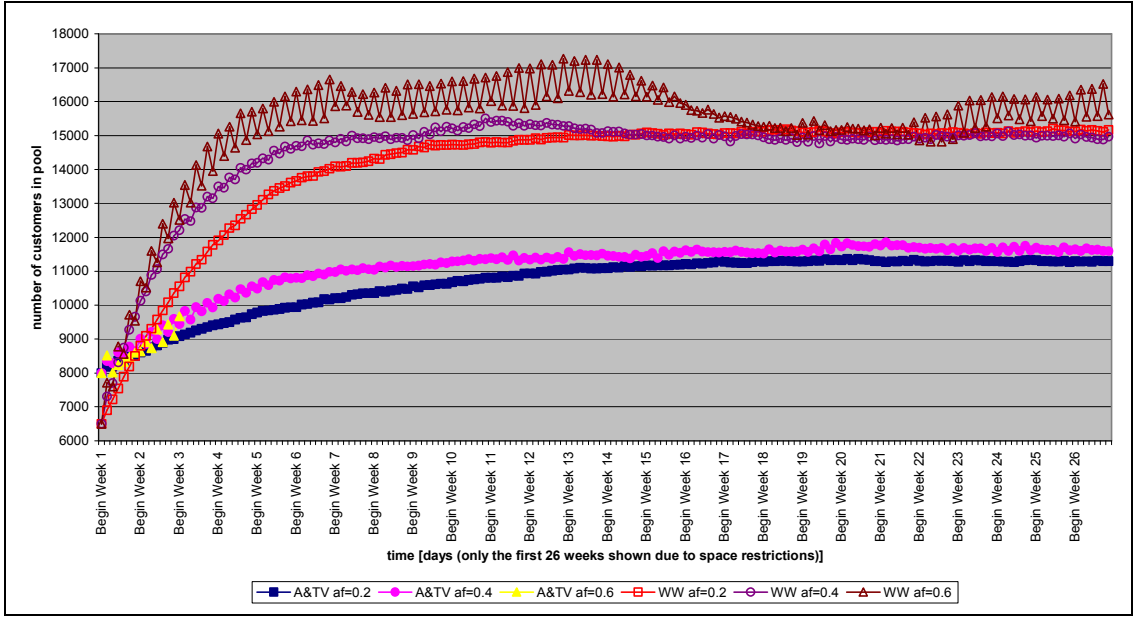

**Figure 12.** Dynamic pool - customer pool size.